\documentclass[amsmath,amssymb,superscriptaddress,nobalancelastpage,aps,prb,twocolumn]{revtex4-1}

\usepackage{graphicx}
\usepackage{varioref}
\usepackage{xr-hyper}
\usepackage{xcolor}
\usepackage{here}
\usepackage{hyperref}
\hypersetup{colorlinks,linkcolor=blue,urlcolor=blue,citecolor=blue}
\usepackage{ulem}
\usepackage{braket}
\edef\restoreparindent{\parindent=\the\parindent\relax}
\usepackage{parskip}
\restoreparindent
\usepackage{gensymb}
\newcommand{\Ca}{Ca$_{2}$RuO$_4$}
\newcommand{\biCa}{Ca$_{3}$Ru$_2$O$_7$}

\newcommand{\Sr}{Sr$_{2}$RuO$_4$}

\newcommand{\tg}{$t_{2g}$}
\newcommand{\dxy}{$d_{xy}$}
\newcommand{\dxz}{$d_{xz}$}
\newcommand{\dyz}{$d_{yz}$}

\begin{document}


\title{Resonant Inelastic X-ray Scattering Study of  Ca$_3$Ru$_2$O$_7$}
    
\author{K. von Arx}
\email{karin.vonarx@uzh.ch}
\affiliation{Physik-Institut, Universit\"{a}t Z\"{u}rich, Winterthurerstrasse 190, CH-8057 Z\"{u}rich, Switzerland} 

\author{F.~Forte}
\affiliation{CNR-SPIN, I-84084 Fisciano, Salerno, Italy}
\affiliation{Dipartimento di Fisica ``E.R.~Caianiello", Universit\`{a} di Salerno, I-84084 Fisciano, Salerno, Italy}

\author{M.~Horio}
\affiliation{Physik-Institut, Universit\"{a}t Z\"{u}rich, Winterthurerstrasse 190, CH-8057 Z\"{u}rich, Switzerland}

\author{V.~Granata}
\affiliation{CNR-SPIN, I-84084 Fisciano, Salerno, Italy}
\affiliation{Dipartimento di Fisica ``E.R.~Caianiello", Universit\`{a} di Salerno, I-84084 Fisciano, Salerno, Italy}

\author{Q. Wang}
\affiliation{Physik-Institut, Universit\"{a}t Z\"{u}rich, Winterthurerstrasse 190, CH-8057 Z\"{u}rich, Switzerland}

\author{L.~Das}
\affiliation{Physik-Institut, Universit\"{a}t Z\"{u}rich, Winterthurerstrasse 190, CH-8057 Z\"{u}rich, Switzerland}
  
\author{Y.~Sassa}
  \affiliation{Department of Physics, Chalmers University of Technology, SE-412 96 Gothenburg, Sweden}
  
\author{R.~Fittipaldi}
\affiliation{CNR-SPIN, I-84084 Fisciano, Salerno, Italy}
\affiliation{Dipartimento di Fisica ``E.R.~Caianiello", Universit\`{a} di Salerno, I-84084 Fisciano, Salerno, Italy}

\author{C.~G.~Fatuzzo}
\altaffiliation{Current address: Materials Sciences Division, Lawrence Berkeley National Laboratory, Berkeley, California 94720, USA  }
\affiliation{Institute of Physics, \'{E}cole Polytechnique Fed\'{e}rale de Lausanne (EPFL), CH-1015 Lausanne, Switzerland}
\affiliation{Physik-Institut, Universit\"{a}t Z\"{u}rich, Winterthurerstrasse 190, CH-8057 Z\"{u}rich, Switzerland}

\author{O.~Ivashko}
\altaffiliation{Current address: Deutsches Elektronen-Synchrotron DESY, 22607 Hamburg, Germany.}
\affiliation{Physik-Institut, Universit\"{a}t Z\"{u}rich, Winterthurerstrasse 190, CH-8057 Z\"{u}rich, Switzerland}

\author{Y. Tseng} 
\affiliation{Swiss Light Source, Photon Science Division, Paul Scherrer Institut, CH-5232 Villigen PSI, Switzerland}

\author{E. Paris} 
\affiliation{Swiss Light Source, Photon Science Division, Paul Scherrer Institut, CH-5232 Villigen PSI, Switzerland}

\author{A.~Vecchione}
\affiliation{CNR-SPIN, I-84084 Fisciano, Salerno, Italy}
\affiliation{Dipartimento di Fisica ``E.R.~Caianiello", Universit\`{a} di Salerno, I-84084 Fisciano, Salerno, Italy}
 
\author{T.~Schmitt}
\affiliation{Swiss Light Source, Photon Science Division, Paul Scherrer Institut, CH-5232 Villigen PSI, Switzerland}

\author{M.~Cuoco}
\affiliation{CNR-SPIN, I-84084 Fisciano, Salerno, Italy}
\affiliation{Dipartimento di Fisica ``E.R.~Caianiello", Universit\`{a} di Salerno, I-84084 Fisciano, Salerno, Italy}

\author{J.~Chang}
\email{johan.chang@physik.uzh.ch}
\affiliation{Physik-Institut, Universit\"{a}t Z\"{u}rich, Winterthurerstrasse 190, CH-8057 Z\"{u}rich, Switzerland}

\begin{abstract}
We present a combined oxygen $K$-edge x-ray absorption spectroscopy (XAS) and resonant inelastic x-ray scattering (RIXS) study of the bilayer ruthenate \biCa. Our RIXS experiments on \biCa\ were carried out on the overlapping planar and inter-planar oxygen resonances, which are distinguishable from the apical one. Comparison to equivalent oxygen $K$-edge spectra recorded on band-Mott insulating \Ca\ is made. In contrast to \Ca\ spectra, which contain excitations linked to Mott physics, \biCa\ spectra feature only intra-$t_{2g}$ ones that do not directly involve the Coulomb energy scale. As found in \Ca, we resolve two intra-$t_{2g}$ excitations in \biCa. Moreover, the lowest lying excitation in \biCa\ shows a significant dispersion, revealing a collective character differently from what is observed in \Ca. Theoretical modelling supports the interpretation of this lowest energy excitation in \biCa\ as a magnetic transverse mode with multi-particle character, whereas the corresponding excitation in \Ca\ is assigned to combined longitudinal and transverse spin modes. These fundamental differences are discussed in terms of the inequivalent magnetic ground-state manifestations in \Ca\ and \biCa.

\end{abstract}

\maketitle

\section{Introduction} 
Transition metal (TM) oxides with 4$d$ valence electrons often exhibit unconventional magnetic and electronic properties. These are dictated by the competition of comparable energy scales set by local interactions, including the Hund's rule and crystal field (CF) terms, together with intrinsic spin-orbit coupling (SOC) of TM ions. By entangling the electron spin to the shape of the electronic cloud in the crystal, SOC makes the electronic spin-orbital states highly sensitive to the inter-site connectivity and effective dimensionality of the underlying lattice. One of the most important consequences is the possibility to tune the relative strength of competing magnetic interactions by varying the effective dimensionality in layered materials.
Calcium-based ruthenates of the Ruddlesden-Popper family Ca$_{n+1}$Ru$_{n}$O$_{3n+1}$ offer one of the richest playgrounds with a great variety of phenomenology. The bilayer compound \biCa\ and its derivatives have been the subject of intense investigations due to a multitude of interesting low-temperature properties, such as spin-valve and giant magnetoresistance effects~\cite{KarpusPRL2004, LinPRL2005,BaoPRL2008,KikugawaJPSJ2010,ZhuPRL2016,XingPRB2018,SowPRL2019,PuggioniPRR2020}.
It has been established that \biCa\ undergoes a magnetic transition at $T_N=56$~K and an electronic transition at $T_s=48$~K~\cite{cao_quantum_2003, yoshida_quasi-two-dimensional_2004,ohmichi_colossal_2004}. The latter transition is -- due to a steep up-rise of the out-of-plane resistivity $\rho_c$~\cite{KikugawaJPSJ2010} -- sometimes referred to as a metal-insulator transition even though the ground state is semi-metallic~\cite{yoshida_quasi-two-dimensional_2004,ohmichi_colossal_2004}. 
Another  reason is the lattice response across $T_s$. Cooling below $T_s$ generates a $c$-axis lattice parameter compression and, through the Poisson's relation, an in-plane lattice parameter enhancement~\cite{yoshidaPRB2005}. This resembles what happens at the metal-insulator transition (350~K) of \Ca~\cite{NakatsujiPRL2004,TakenakaNatComm2017}. There, the $c$-axis compression leads to an almost fully occupied \dxy\ orbital and a Mott-gap opening in the half-filled \dxz\ and \dyz\ bands. However, the effect in \biCa\ is much smaller (0.1\% and $>$1\% compression of the lattice parameter $c$ in \biCa\ and \Ca, respectively)~\cite{yoshidaPRB2005,friedt_structural_2001}.
The fact that both \biCa\ and \Ca\ undergo similar $c$-axis compressive transitions but end up with different ground states makes comparative studies interesting. In addition to the electronic properties, the magnetic ground states of these two compounds differ as well. Whereas \Ca\ displays a G-type antiferromagnetic state below $T_N=110$~K~\cite{braden_crystal_1998}, the in-plane magnetic moments in \biCa\ order ferromagnetically, leading to an A-type antiferromagnetic state~\cite{yoshidaPRB2005}. This difference in the in-plane magnetic order implies that the interaction within the layers plays an important role for the magnetic ground state of these compounds. The investigation of the magnetic and orbital degrees of freedom and their excitation spectrum therefore offers a view on the complex interplay between different energy scales relevant for the ground state. In this respect, recent spectroscopic and neutron scattering measurements demonstrated that the magnetic ordering in \Ca\ may sustain both longitudinal and transverse magnon modes with a large anisotropy gap, which reflects the impact of broken tetragonal symmetry in combination with SOC~\cite{jain2017higgs, souliou_raman_2017,das_spin-orbital_2018}.\par
\begin{figure*}[t]
	\begin{center}
		\includegraphics[width=0.999\textwidth]{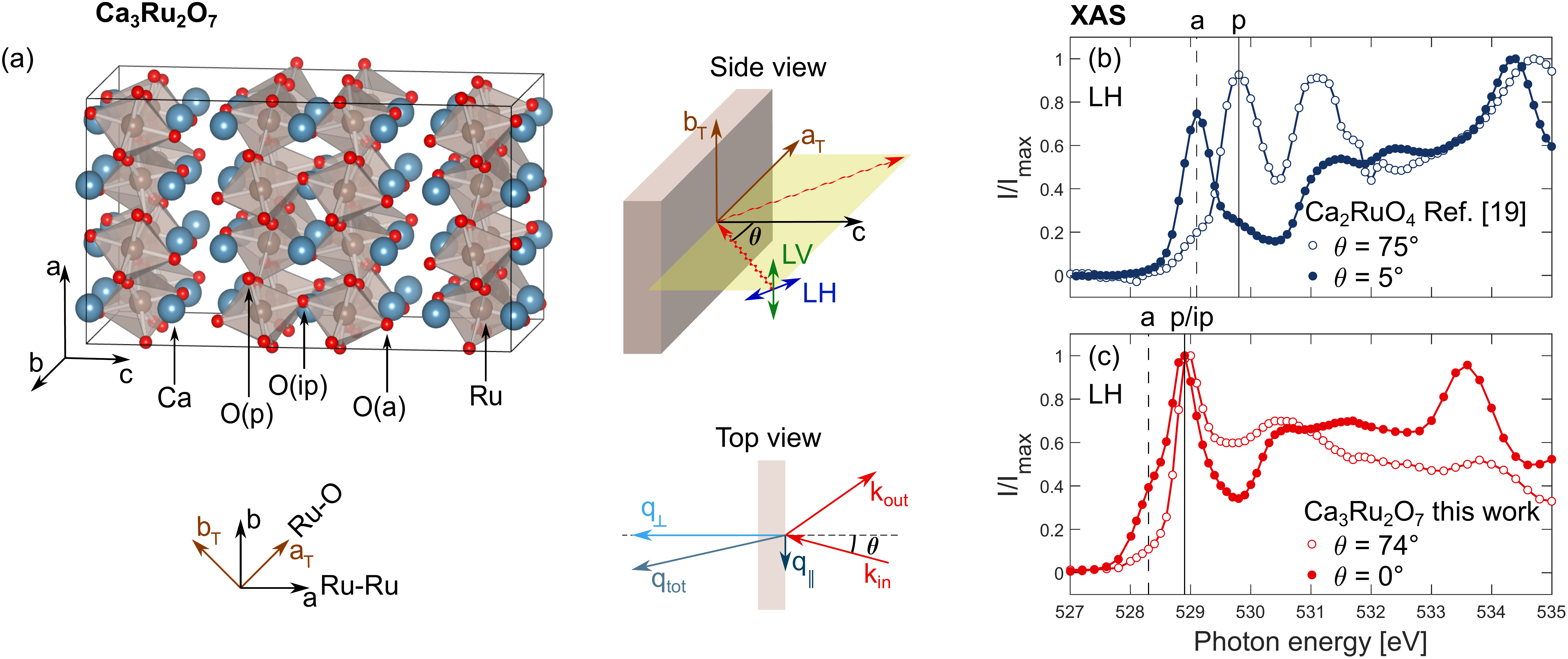}
	\end{center}
	\caption{(a) Schematically depicted RIXS geometry with respect to the crystal lattice of \biCa. Different oxygen sites in the crystal structure are labelled as O(p), O(a) and O(ip) for planar, apical and inter-planar sites, respectively. Momentum dependence was measured along the Ru-O direction. Therefore, the reciprocal space is indexed in tetragonal notation with $a_T\approx b_T\approx1/2\sqrt{a^2+b^2}$, where $a=5.37$ {\AA} and $b=5.54$ {\AA}~\cite{yoshidaPRB2005}. (b), (c) Background subtracted and normalized XAS spectra of \Ca\ and \biCa, respectively. The spectra were recorded with LH polarization near grazing and normal incident light directions as indicated. The dashed vertical black line indicates apical, whereas the solid vertical black line indicates planar (\Ca) and overlapping planar and inter-planar (\biCa) oxygen resonances probing Ru $t_{2g}$ states. Dark blue data on \Ca\ are taken from ref.~\onlinecite{das_spin-orbital_2018}.}
	\label{fig:fig1}
\end{figure*}
In this paper, we present a combined  oxygen $K$-edge x-ray absorption spectroscopy (XAS) and resonant inelastic x-ray scattering (RIXS) study of \biCa\ and compare it to previously published work on \Ca~\cite{FatuzzoPRB2015,das_spin-orbital_2018}, with the aim to investigate the distinctive fingerprints of the magnetic state in the single and bilayer compounds. With this methodology, the Ru 4$d$ orbitals are accessed indirectly through their hybridization with oxygen $p$ orbitals. In this fashion, we probe the two unoccupied \tg\ states.  This indirect approach has routinely been applied to different TM oxides~\cite{bisogni_bimagnon_2012,salaPRB2014,lu_dispersive_2018,
pinciniPRB2019}.\par
Our study demonstrates that in \biCa, only the two lowest intra-$t_{2g}$ excitations are observed, whereas in \Ca, the Mott insulating ground state produces a set of excitations within the $t_{2g}$ subspace, which consists of two low-energy and two mid/high-energy structures.
An important difference -- the main observation reported here -- is that the lowest lying excitation exhibits a clear dispersive character in \biCa. This marked collective behavior is not found with the corresponding excitation in \Ca. The fundamentally different magnetic ground states of \Ca\ and \biCa\ are therefore manifested in the excitation spectrum, both within the $t_{2g}$ and between the $t_{2g}$ and $e_g$ sectors. We discuss this within the theoretical framework of fast collision approximation for the RIXS cross-section~\cite{AmentPRB2007,amentRMP2011}. Taking into account the different magnetic ground states of \Ca\ and \biCa, qualitative agreement between calculated and observed RIXS spectra is obtained. The model qualitatively describes the marked differences in the RIXS spectra recorded on \Ca\ and \biCa.
Moreover, we analyze the nature of the lowest lying intra-$t_{2g}$  excitation. In this fashion, we show that this excitation is magnetic in both compounds, but with fundamentally different natures. In \Ca, the lowest lying excitation is consistent with composite longitudinal amplitude and transverse spin modes, whereas in \biCa\ it has a dominant transverse spin nature.\par
These results provide decisive evidence for the capability of oxygen $K$-edge RIXS in probing the complex structure of electronic excitations in 4$d$ ruthenates. Particularly, it is confirmed that the low-energy spin/orbital modes are also directly accessible in virtue of modest SOC~\cite{lu_dispersive_2018}. Such elementary excitations reflect the balance among competing interactions, being therefore crucial for revealing the origin of emergent phases and for determining the low-energy Hamiltonian in layered ruthenates, where magnetic interactions are no longer dictated by a global spin SU(2) symmetry alone.
\begin{figure*}[t]
	\begin{center}
		\includegraphics[width=0.999\textwidth]{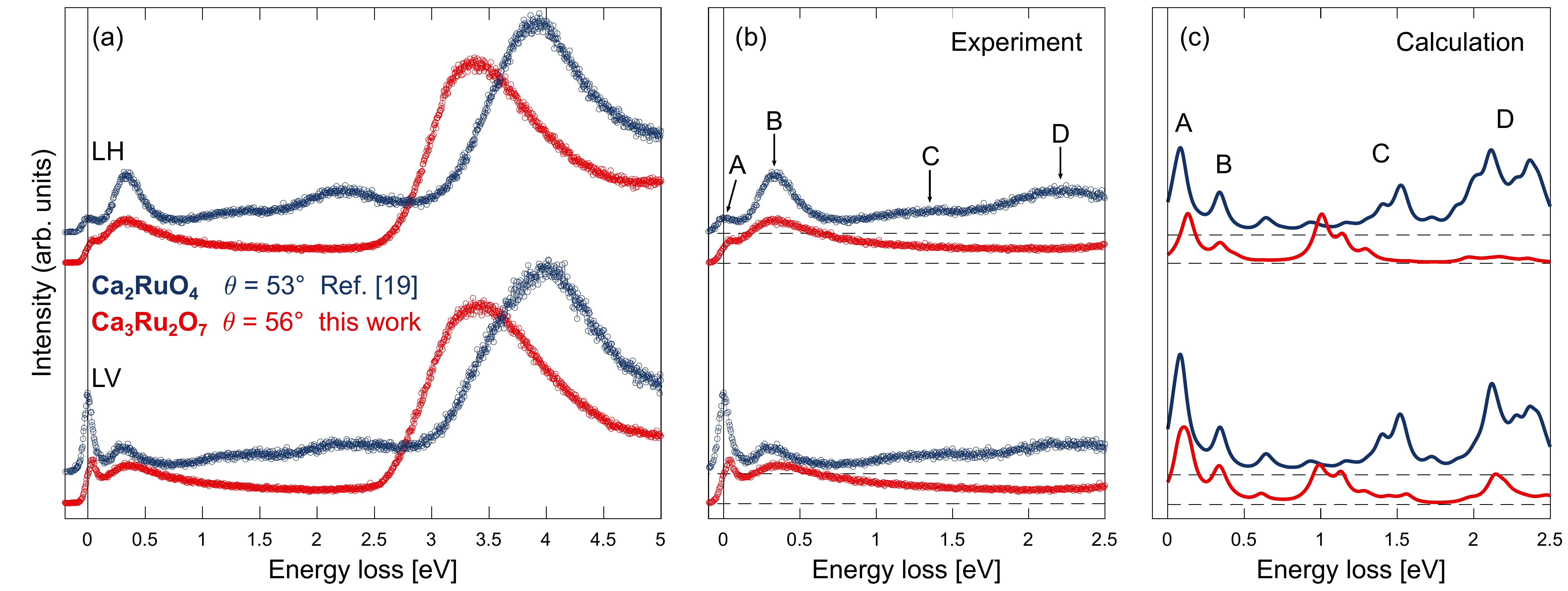}
	\end{center}
	\caption{(a) RIXS spectra of \Ca\ and \biCa\ recorded with LH and LV polarization as indicated. To enhance visibility, the spectra are given an individual vertical shift. Dark blue data on \Ca\ are taken from ref.~\onlinecite{das_spin-orbital_2018}. (b) Zoom of (a) to show the low-energy excitations labelled as A, B, C and D. (c) Calculated RIXS spectra of \Ca\ and \biCa\ for LH and LV polarization.}
	\label{fig:fig2new}
\end{figure*}

\section{Methods} 
High quality single crystals of \biCa\ were grown by the floating zone techniques~\cite{FukazawaPhysB00,snakatsujiJSSCHEM2001}, aligned \textit{ex-situ} by x-ray Laue and cleaved \textit{in-situ} using the top-post method. XAS and RIXS~\cite{amentRMP2011} measurements were carried out at the ADRESS beamline~\cite{ghiringhelliREVSCIINS2006, strocov2010high} of the Swiss Light Source (SLS) at the Paul Scherrer Institut. The scattering geometry is indicated in Fig.~\ref{fig:fig1}(a). A fixed angle of 130$^\circ$ between incident and scattered light was used. In-plane momentum ${\bf{q}}_{||}=(h \, 2\pi/a,k \, 2\pi/b)$ is varied by controlling the incident photon angle $\theta$. In this work, the reciprocal space is indexed in tetragonal notation. 
Grazing and normal incidence conditions refer to $\theta\approx 90^\circ$ and $0^\circ$, respectively.
Linear vertical (LV) and horizontal (LH) light polarizations were used to probe the oxygen $K$-edge at which an energy resolution
of 22.5~meV (Gaussian standard deviation $\sigma$) on \biCa\ spectra was obtained. Elastic scattering is modelled by a Gaussian lineshape (see Appendix C for details) with $\sigma$ set by the energy resolution. The presented data on \biCa\ is collected at the base temperature $T=20$~K unless otherwise indicated. Our experimental setup for XAS and RIXS measurements on \biCa\ is equivalent to that previously used for measurements of \Ca. Additionally, the energy resolution is comparable in both experiments and the minor base temperature differences (16 - 20 K) are negligible compared to the magnetic and electronic transitions in the two systems. Therefore, results on the two compounds are directly comparable.

\section{Results} 
The oxygen $K$-edge XAS spectra taken with LH light polarization on \Ca\ and \biCa\ are shown in Figs.~\ref{fig:fig1}(b) and (c). For \Ca, the apical and planar oxygen resonances are disentangled by using LH light near normal or grazing conditions respectively~\cite{FatuzzoPRB2015,das_spin-orbital_2018}. 
For tetragonal and their orthorhombic derivatives, the crystal field (chemical) environment\cite{chen_out--plane_1992,NohPRB2005,salaPRB2014} and  Coulomb interaction impose the apical oxygen edge to appear at lower photon energy than the planar oxygen edge. On this basis,
the first and second pre-edges, indicated by dashed and solid vertical lines, correspond respectively to the resonances at the apical and planar oxygen sites, from which hybridization with Ru $t_{2g}$ orbitals takes place.
In the case of \biCa, there are three oxygen sites: 
planar O(p), inter-planar O(ip) and apical O(a)
-- see Fig.~\ref{fig:fig1}(a). Compared to \Ca, these oxygen sites are harder to distinguish in the XAS spectra of \biCa\ as the planar O(p) and inter-planar O(ip) 
sites have similar CF environments.
Similarly to other layered oxides~\cite{salaPRB2014,chen_out--plane_1992}, we assign the first pre-edge of \biCa\ to the O(a) site.  In the normal  ($\theta$ = 0$^{\circ}$) condition, this resonance appears as a shoulder (528.3~eV) on the second pre-edge (528.9~eV) that is assigned to the O(p) and O(ip) sites. The reduced splitting of the oxygen $K$-pre-edges is also known from the XAS study of the Ruddlesden-Popper Sr$_{1+n}$Ru$_n$O$_{3n+1}$ series with $n=1,2,$ and 3~\cite{malvestutoPRB2011, malvestutoPRB2013}. Eventually, for cubic SrRuO$_3$, the two pre-edges merge together and only one feature is observed~\cite{guedes_core_2012}. The features at higher energies correspond to resonances probing the O $p$ orbitals hybridized with the unoccupied Ru $e_{g}$ states.\par 

\begin{figure*}[t]
	\begin{center}
		\includegraphics[width=0.999\textwidth]{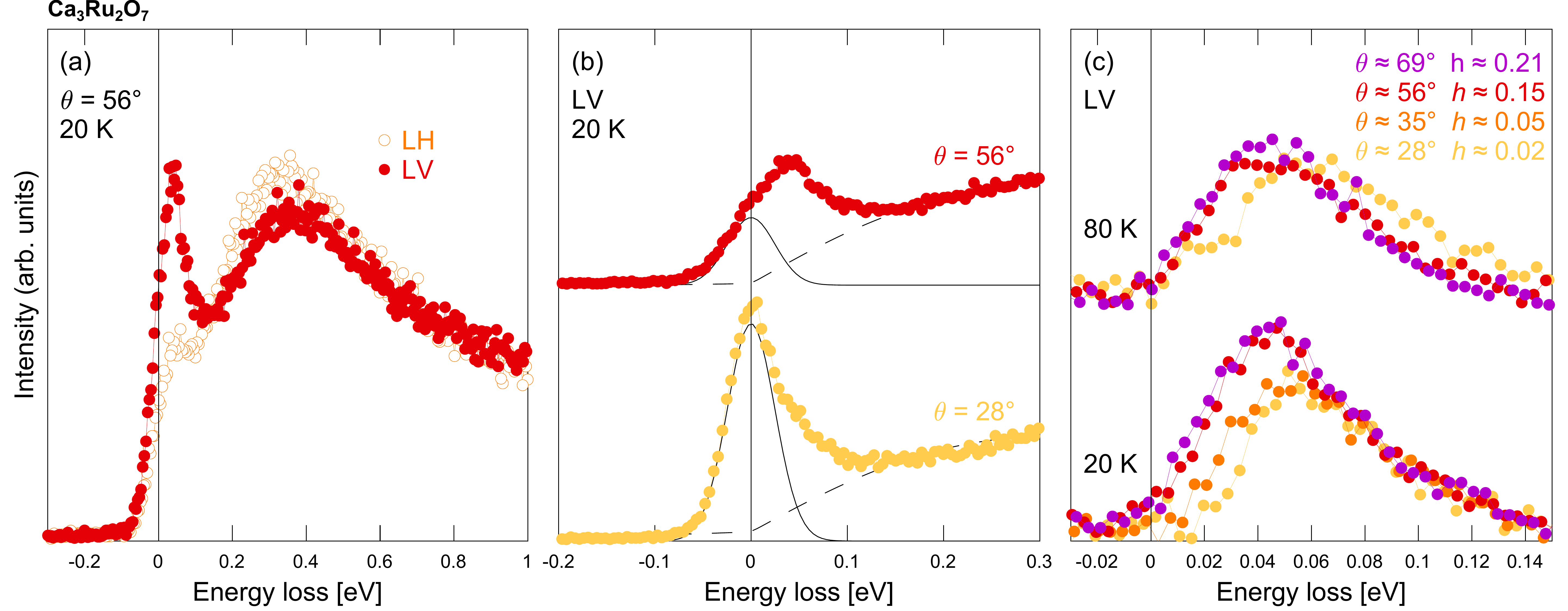}
	\end{center}
	\caption{RIXS spectra of \biCa\ focusing on the low-energy part. (a) Comparison between LH and LV spectra at the same incidence angle. (b) Low-energy part for two different incidence angles measured with LV polarization. Solid black lines are Gaussian fits with a width fixed to the energy resolution of the experiment to model the elastic scattering. Dashed black lines are the sums of B excitation and background contributions. To enhance visibility, the spectrum at higher angle is given a vertical shift. (c) RIXS spectra after subtraction of the elastic line, B excitation and background contributions to show the dispersion of the A excitation at 20~K and 80~K, as indicated. To enhance visibility, the spectra at 80~K are given a vertical shift. The indicated $\theta$ and $h$ values correspond to the 20~K data, values for 80~K data differ slightly -- see Fig.~\ref{fig:fig4}.}
	\label{fig:fig3}
\end{figure*}

For our RIXS study of \biCa, we have focused entirely on the most intense oxygen $K$-pre-edge (528.9~eV) that probes the planar and inter-planar sites. In Figs.~\ref{fig:fig2new}(a) and (b), spectra recorded with LV and LH light are compared to the corresponding planar spectra of \Ca.
First, we notice that the "block" of $dd$-excitations in \biCa\ around 3.5~eV is consistently shifted to lower energies relatively to what is found in \Ca. Another noticeable difference is that among the four "low" energy excitations reported~\cite{das_spin-orbital_2018} for \Ca\ [labelled as A,B,C, and D in Fig.~\ref{fig:fig2new}(b)], only the two lowest (A and B) are found in \biCa. 
The B excitation of bilayer \biCa\ has a significantly smaller amplitude and is much broader than in \Ca. However, for both compounds, the B excitation is more intense when probed with LH polarization, see Fig.~\ref{fig:fig3}(a). \par
The lowest lying excitation (labelled as A) is overlapping with the elastic line and careful analysis is required to separate these two contributions. Elastic scattering is most pronounced near the specular condition, therefore the A excitation appears as a shoulder on the energy loss side -- see Fig.~\ref{fig:fig3}(b).
Near grazing condition, the situation is reversed and the elastic scattering appears as a shoulder on the left side of the A excitation peak.
To model the elastic contribution, we use a Gaussian profile with the linewidth set by the energy resolution. In this fashion, it is possible to extract the A excitation by subtracting the elastic component as well as the contributions from the B excitation and background, as illustrated in Figs.~\ref{fig:fig3}(b) and (c).
As the incidence angle -- and hence the in-plane momentum transfer -- is varied, 
the A excitation is dispersing to a lower energy away from the zone center. Finally, the A excitation persists at least up to 80 K, as shown in Fig.~\ref{fig:fig3}(c).\par

The momentum dependence of the A and B excitations extracted from the \biCa\ data are compared in Fig.~\ref{fig:fig4} to the corresponding excitations in \Ca. For the B excitation, the peak position is defined as the maximum obtained from the derivative of the spectrum, since the peak is extremely broad. Within the energy resolution of this experiment, no momentum dependence can be resolved for this excitation in \biCa. 
The situation is different in \Ca, where a small upward dispersion away from the zone center is detected for the B excitation~\cite{das_spin-orbital_2018}.
Most pronounced differences are observed for the A excitation. The strong dispersion found in \biCa\ is completely absent in \Ca. Additionally, the excitation is located at significantly higher energies in \Ca\ at around 80~meV compared to 55~meV in \biCa. We stress that the A excitation dispersion is measured with LV polarization and 
hence is probed on either the planar or inter-planar site via the $p_y$ orbital independently of $\theta$. The observed dispersion can therefore not be assigned to scattering geometry effects.

\section{Discussion}
To discuss the XAS spectra, we first summarize the interpretation of the \Ca\ data published recently~\cite{FatuzzoPRB2015,das_spin-orbital_2018}. The exact mechanism behind the Mott insulating state of \Ca\ has long been under discussion and various theoretical models have been proposed~\cite{KogaPRL2004,LiebschPRL2007,gorelovPRL2010}. In this context, the \Ca\ XAS results strongly support the explanation via a complete orbital polarization with the almost fully occupied $d_{xy}$ orbital. 
Indeed, the XAS spectra, reproduced in Fig.\ref{fig:fig1}(b), are in perfect accordance with this picture. As discussed in ref.~\onlinecite{das_spin-orbital_2018}, the dominant XAS response flips from the apical to the planar resonance when changing from normal to near grazing incidence using LH polarized light. This geometry effect is a result of the almost fully occupied Ru $d_{xy}$ orbital, that is unavailable for absorption.
Comparing the \Ca\ and \biCa\ spectra, the differences in crystal structure and orbital occupation become apparent. Due to the nonequivalent apical oxygen sites in \biCa, the apical feature splits and the outer apical O(a) is only visible as a shoulder to the strong planar resonance that overlaps with the inter-planar O(ip) resonance. Taking into account the relative intensities of the two features, the XAS results suggest a different orbital occupation than in \Ca, with an only partially filled $d_{xy}$. This partial occupation is also in accordance with the reduced $c$-axis compression in \biCa\ compared to \Ca~\cite{yoshidaPRB2005,friedt_structural_2001}. \par
\begin{figure}
	\begin{center}
		\includegraphics[width=0.48\textwidth]{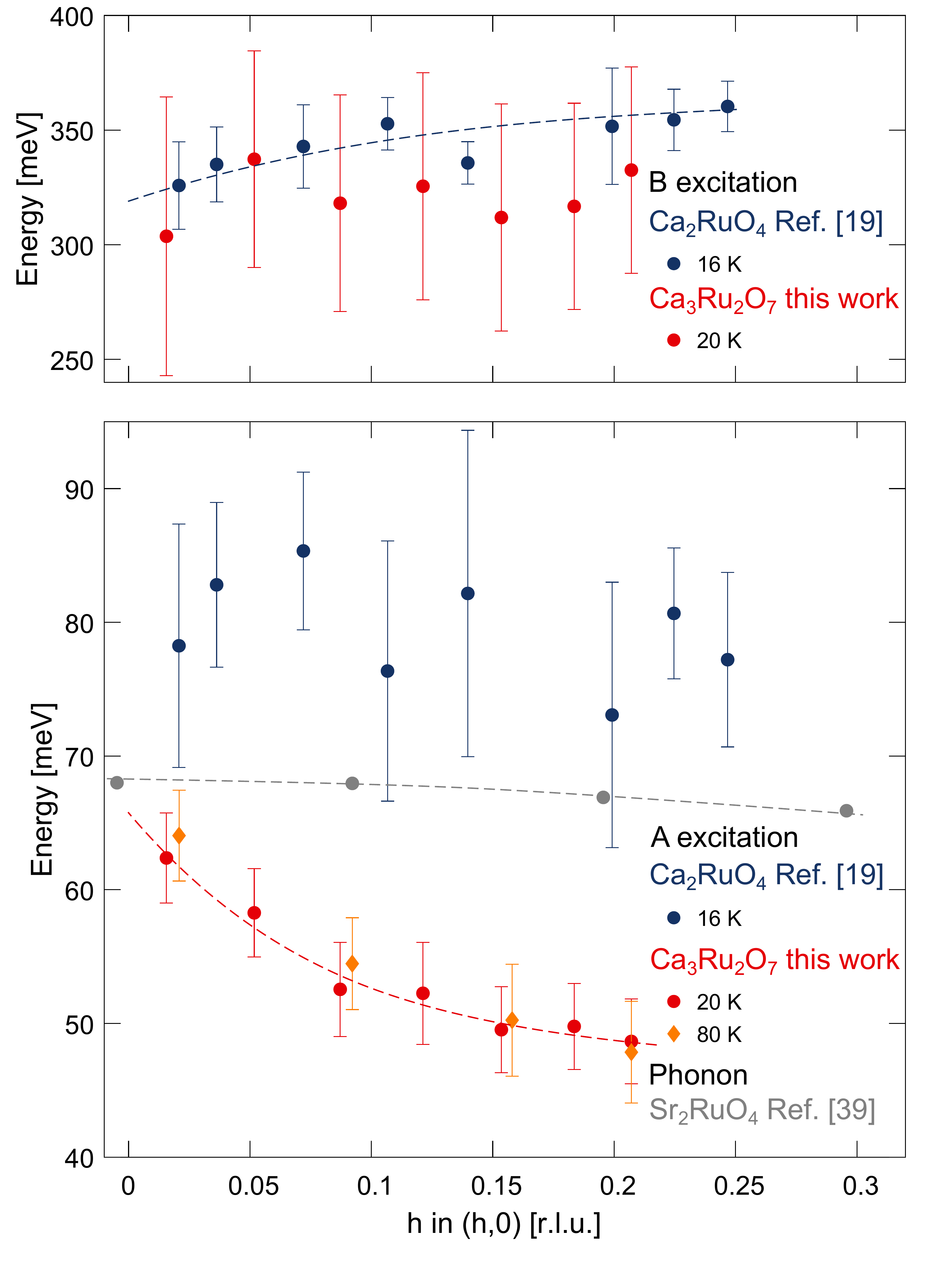}
	\end{center}
	\caption{Dispersion of the low-energy excitations along $\theta$ which corresponds to a varying in-plane momentum ${\bf{q}}=(h,0)$ for the A (bottom) and B (top) excitations. Error bars for the A excitation in \biCa\ indicate standard deviation $3\sigma$ of the fitted peak position. By contrast, for the B excitation, the error bars reflect the peak width at 98\% of the total height. Dark blue data show the dispersion for the single layer \Ca, taken from ref.~\onlinecite{das_spin-orbital_2018} including error bars. Grey data show a phonon dispersion in \Sr\ measured by inelastic neutron scattering~\cite{braden_lattice_nodate}. Dashed lines are guides to the eye.}
	\label{fig:fig4}
\end{figure}

Next, we turn to discuss the RIXS spectra.
The fact that completely different oxygen $K$-edge RIXS spectra are observed for \Ca\ and \biCa\ is a beautiful example of how ground state fingerprints are encoded into the excitations. In principle, the CF environment around an in-plane oxygen should be similar for \Ca\ and \biCa. Yet, the RIXS excitation spectra are fundamentally different for these two compounds. In \Ca, a sequence of excitations has been identified in the $t_{2g}$ sector, which are separated from the higher energy $t_{2g}\rightarrow e_g$ features in the energy range $\sim$ 3--5 eV~\cite{das_spin-orbital_2018,GretarssonPRB2019} -- see Fig.~\ref{fig:fig2new}(a). In particular, two broad excitations located around 1 eV and 2 eV, labelled as C and D, are linked to the energy scales of Hund's coupling and Coulomb interaction responsible for the Mott insulating ground state. In semi-metallic \biCa\ by contrast, these excitations are completely absent -- see Fig.~\ref{fig:fig2new}(b). Even within the lower energy $t_{2g}$ sector, pronounced differences are identified. Although two excitations (labelled as A and B) -- with similar energy scales -- are resolved for both compounds, they appear to have a fundamentally different nature. In \biCa\ the lowest lying excitation is clearly dispersive, whereas in \Ca\ no dispersion was resolved for the corresponding branch.\par
To gain insight into the microscopic picture behind these excitations, the RIXS response was modelled for both compounds, and compared to the experimental spectra in Fig.~\ref{fig:fig2new}. We used the fast collision approximation~\cite{AmentPRB2007,amentRMP2011} of the RIXS cross-section describing the light-induced excitation -- and subsequent absorption -- of an electron from the O $1s$ level into the $2p$ level, for both LV and LH incoming polarization. Full detailed description of this approach is reported in Appendices A and B. The RIXS intensity  was calculated  via exact diagonalization of a model Hamiltonian defined on a cluster of two ruthenium sites connected by one planar oxygen site along an in-plane direction. The bond bending due to the rotation of the octahedra around the $c$ axis is allowed. The ruthenium-site Hamiltonian is defined on the $t_{2g}$ subspace and consists of three terms: (1) CF splitting $\Delta$ between the $d_{xy}$  and  $d_{xz}$, $d_{yz}$ orbitals, 
(2)  SOC $\lambda$, and
(3)  Coulomb interaction, which is expanded into intra-orbital and inter-orbital Hubbard interactions of strengths $U$ and $(U-5J_{\mathrm{H}}/2)$, respectively. Inter-orbital Hund's coupling as well as the pair-hopping term, are both of strength $J_{\mathrm{H}}$. Material specific
values $\Delta = 0.3$~eV, $\lambda=0.05$~eV, $U=2.0$~eV, and $J_{\mathrm{H}}=0.4$~eV~\cite{mizokawaPRL2001,veenstraPRL2014,FatuzzoPRB2015,han_lattice_2018}, are used to evaluate the model for both \Ca\ and \biCa. Similar values of $\Delta$, $U$ and $J_{\mathrm{H}}$ have been used for DMFT calculations~\cite{SutterNatComm2017a} of \Ca\ and are comparable to those used in modelling the spin-excitation dispersion observed by neutron scattering~\cite{jain2017higgs}, RIXS spectra~\cite{das_spin-orbital_2018}, as well as magnetic anisotropy~\cite{porterPRB2018}. Here, we point out that small differences from previous estimates of the microscopic parameters are fully awaited since, in our description, the oxygen degrees of freedom are explicitly included, and this may lead to a renormalization of the local interaction terms. To take into account the different ground states, \Ca\ is modelled with an antiferromagnetic (AFM) in-plane interaction, whereas we consider an extra exchange field to stabilize the ferromagnetic (FM) ground state and spins along the in-plane easy axis in \biCa~\cite{braden_crystal_1998, yoshidaPRB2005}. Henceforth, we will also refer to the \Ca\ and \biCa\ bonds as AFM and FM, respectively, corresponding to the G-type and A-type AFM structures.\par
In Fig.~\ref{fig:fig2new}(c) the calculated RIXS responses in LH and LV polarizations are presented for both ground states, showing a reasonable 
overall agreement with the experimental spectra in Fig.~\ref{fig:fig2new}(b). In both cases four distinct excitations are evident,  with approximate energy losses of 0.08, 0.4, 1--1.5 and 2 eV. In the FM  case, we observe an overall decrease of the peak intensities. This effect is even more pronounced for excitations above 1~eV.\par
 \begin{figure}
	\begin{center}
		\includegraphics[width=0.48\textwidth]{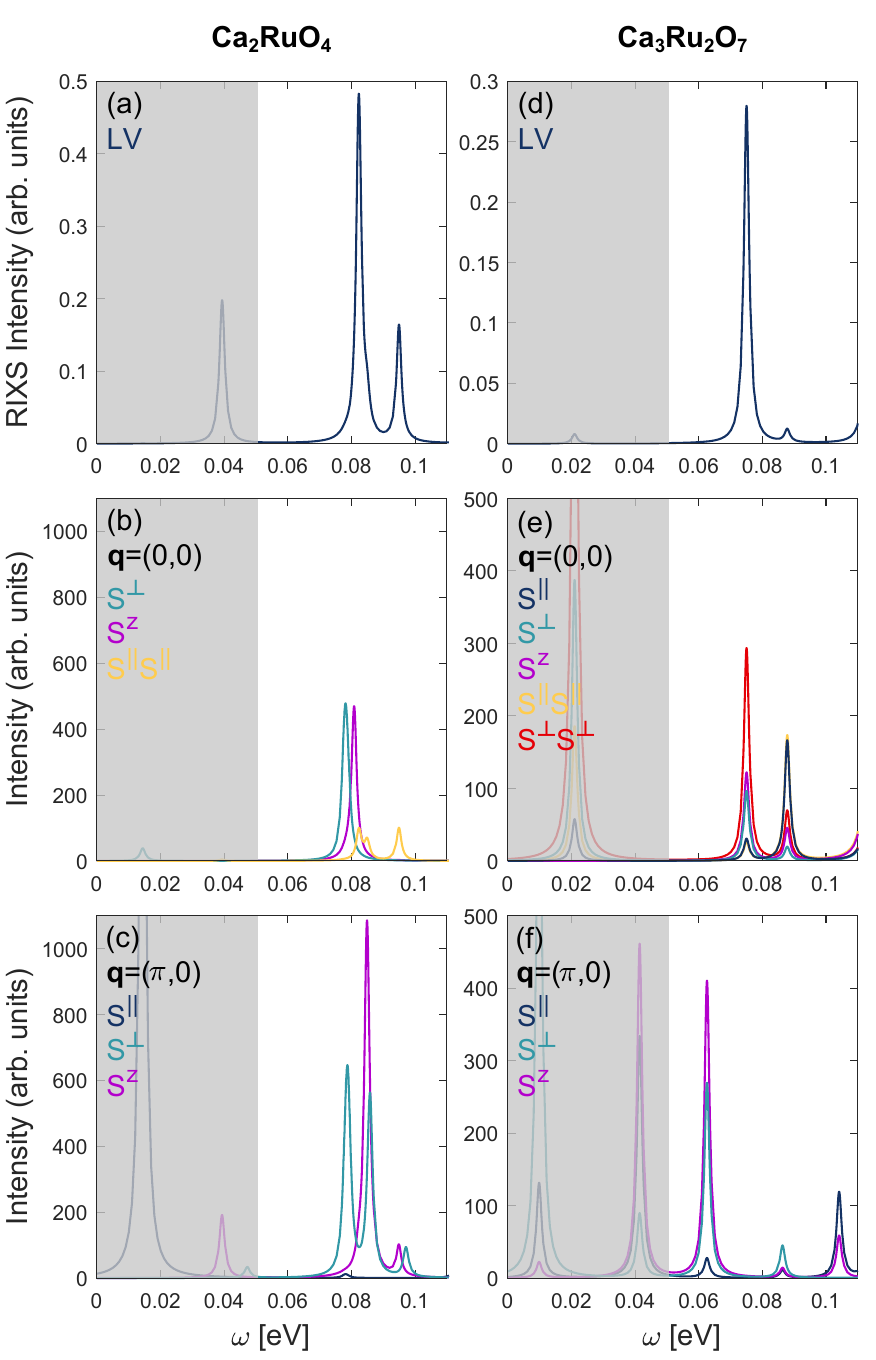}
	\end{center}
	\caption{Theoretical model calculation of the RIXS spectra in panels (a) and (d) and the dynamic spin structure factors $S^{\mu}({\bf q},\omega)$, and spin-spin dynamic spin structure factors  $(S_{i} \cdot S_{i+1})^{\mu}( {\bf q}, \omega)$, $\mu=||, \perp, z$, for ${\bf q}=(0,0)$ and $(\pi,0)$ in the low energy regime for \Ca\ and \biCa. The gray shaded area corresponds to the energy region currently not accessible by RIXS due to the limited resolution.}
	\label{fig:fig5}
\end{figure}
The origin of  the four features for \Ca\ has already been assigned in a previous work, and we recall it here for convenience~\cite{das_spin-orbital_2018}. We point out  that, in the present simulation, the RIXS intensity has been evaluated by fully taking into account the scattering geometry in Fig.~\ref{fig:fig1}(a), and that the LV and LH spectra have been obtained by averaging the spectra over two orthogonal in-plane bond directions of the cluster.\par
Excitations A, B, C, and D in \Ca\ have been interpreted on the basis of the multiplet structure of the $d^4$ configuration of the Ru$^{4+} $ ion. In particular, they are associated to transitions within low-energy spin-orbital configurations, which have one doubly occupied orbital (doublon), or two doubly occupied orbitals. In the framework of the ionic picture, structures C and D have been assigned to $J_H$ driven spin-state transitions between $S=1$ and $S=0$ states in the single- and two-doublon sectors~\cite{das_spin-orbital_2018}. The partial suppression of the weight associated to the features above 1 eV in the case of a FM configuration is a consequence of the Pauli blocking of those intra-$t_{2g}$ transitions. This mechanism may justify the lowered intensity of C and D structures in the FM background of Ru-O planes in \biCa. We notice that the experimental results suggest a stronger suppression than predicted by the model calculations. A reason for the discrepancies between experiment and the modelling may have to do with the less insulating ground state of \biCa. The cluster calculations may not capture precisely the more delocalized nature of \biCa.\par
In \Ca, the lowest energy features A and B are associated to spin-orbital excitations within the $S=1$ subspace of the 
$t_{2g}$ multiplets, whose energies are determined by the relative strength of the CF potential and the SOC. Even though they are not fully resolved experimentally due to limited energy resolution, those excitations are accessible in oxygen $K$-edge because of the SOC in the Ru $4d$ shell which strongly hybridizes $4d$ states with O $2p$ orbitals. In particular, the B structure has been attributed to multiple transitions to the highest energy $S=1$ spin-orbital sector, while the A structure has been generically associated to composite magnetic transitions within the lowest-energy sector. We observe that, in the FM case, features occurring at a similar energy scale are observed, and we want to elucidate the possible spin-orbital (magnetic) origin of these excitations, with a special focus on the lowest A feature.\par
Beforehand, we observe that the lowest lying excitation A has an energy scale typical of both optical phonons and magnons. The strong dispersion of this excitation near the zone center is, however, atypical for optical phonons. In Sr$_2$RuO$_4$, where complete phonon dispersions have been calculated and probed by neutron scattering~\cite{braden_lattice_nodate,wang_phonon_nodate}, optical phonons are in fact found at $\sim70$ and $\sim90$~meV. None of them has a dispersion around the zone center compatible with what we observe in \biCa. This is demonstrated in Fig.~\ref{fig:fig4}, where the relevant optical Sr$_2$RuO$_4$ phonon dispersion -- measured by neutron scattering -- is shown in grey.  In comparison to the A excitation measured in \biCa, this optical phonon is non-dispersive. Moreover, the 70 meV optical phonon found in Sr$_2$RuO$_4$ stems from vibrations of the apical oxygen whereas we are probing on the planar and inter-planar 
oxygen sites. On this basis, assigning a magnetic origin to the lowest lying excitation appears the most plausible interpretation. We stress that it is not unusual to observe magnetic excitations beyond the magnetic ordered state due to the persistence of short-range magnetic correlations. In cuprates and iron pnictides, paramagnon excitations are found deep into the magnetically disordered state~\cite{DeanNatMat2013, monney_resonant_2016,pelliciariComPhys2019}. Observing no significant temperature dependence of the A excitation dispersion is therefore expected.\par
To further verify the magnetic origin of the A excitations in \Ca\ and \biCa, and to reveal their distinct nature, we evaluated the dynamic spin structure factors $S^{\mu}({\bf q},\omega)$, and spin-spin dynamic spin structure factors  $(S_{i} \cdot S_{i+1})^{\mu}( {\bf q}, \omega)$, $\mu=x, y, z$, [Fig.~\ref{fig:fig5}(b),(c),(e),(f)] for ${\bf q}=(0,0)$ and $(\pi,0)$, which are the only viable values for the momentum transfer of our Ru-O-Ru cluster. Here, we point out that the ground state is made by magnetic moments that are aligned in the Ru plane. For convenience, we refer to $||, \perp,z$, for a spin mode excitation that is collinear, perpendicular in-plane, perpendicular out-of-plane with respect to the orientation of the ordered magnetic moments in the ground state, respectively.\par
Let us start with the AFM case in Fig.~\ref{fig:fig5}(a)--(c). The comparison of the low-energy part of the RIXS spectrum for the AFM configuration with the calculated spin structure factors at ${\bf{q}}=(0,0)$ allows to associate the dominant excitation in the RIXS spectrum to features with transverse single spin modes, i.e. $S^z$ and $S^{\perp}$, and longitudinal two-spin $S^{||}S^{||}$ correlation functions. 
This is consistent with the previous interpretation of the A excitation as evidence of composite excitations such as  longitudinal (Higgs) two-particle and transverse bimagnon modes~\cite{souliou_raman_2017,das_spin-orbital_2018}. In particular, from the analysis we observe that the lowest energy spin excitations occur at about 20 and 40 meV -- see Fig.~\ref{fig:fig5}(b),(c), mainly through single spin flip at each Ru site.The former energy scale is related to the effective single ion anisotropy due to the interplay of spin-orbit and crystal field potential. A distinctive aspect of the magnetic ground state is that, due to the spin-orbit coupling and crystal field potential, there is neither rotational nor parity conservation for the local spin. The resulting ground state is then a quantum superposition of several components. Specifically, it consists of dominant exchange driven anisotropic antiferromagnetic correlations, and it also includes states corresponding to the variation of amplitude and direction of the local $S=1$ magnetic moments with respect to the easy axis. This peculiar character of the ground state allows to have a significant spectral weight associated to high-energy excitations, corresponding to the RIXS active states close to 80 meV -- see Fig.~\ref{fig:fig5}(a). Taking into account the energy profile of the dynamic spin response, we deduce that the modes at about 80 meV have a multi-particle spin character, as they are accessible by means of both single transverse and double longitudinal spin excitations. Our results also predict the existence of a lower energy feature in the RIXS spectrum, having similar character, at 40 meV, in an energy range which is not detectable in the present experiment.\par
The FM case offers a similar result, since the lowest A feature may also be associated to magnetic excitations. Notably in this case, the dominant excitation occurs at slightly lower energy, and corresponds to mainly transverse spin excitations, of single- $S^{z}$ and $S^{\perp}$ or two-particle type $S^{\perp}S^{\perp}$. Moreover, according to the simulation, the existence of a very weak feature located at 20 meV is also predicted.\par
Having identified the nature of the magnetic excitations associated to the lowest RIXS feature in both the AFM and FM ground states, one can also estimate the bandwidth of the continuum of the corresponding collective modes propagating along the $(0,0) \rightarrow (\pi,0)$ path. Comparing the relevant $S^{\mu}({\bf q},\omega)$ at ${\bf q}=(\pi,0)$ and $(0,0)$ shows that, in the AFM case, magnetic peaks are located approximately in the same energy range at different wave vectors. On the contrary, in the FM configuration, the peaks associated to the single spin excitations are shifted to lower energies by 20-30 meV, when going from ${\bf q}=(0,0)$ to ${\bf q}=(\pi,0)$. This is in accordance with what is observed in the experimental spectra -- see Fig.~\ref{fig:fig4}. We also carried out the calculation of the local and two-site orbital angular momentum correlation functions, which reveal that the A peak in the AFM case has significant orbital contribution, while it is substantially suppressed in the FM case. This is consistent with the observation that the FM ground state has a different orbital pattern~\cite{fortePRB2010} when compared to the AFM configuration. The doublon can have a stronger tendency to occupy different orbitals on neighbouring Ru sites in the FM case. Moreover, the spin-orbit coupling tends to align the orbital moments; since the Ru spins are also ferromagnetically correlated. This implies that orbital variations can be suppressed in the low-energy spin sector. Here, we argue that the lack of the orbital component in the targeted excitation allows to have a larger effective exchange, which results in an enhancement of the bandwidth as we find in the cluster analysis.

\section{Conclusions}
In summary, we have carried out a combined oxygen $K$-edge XAS and RIXS study of \biCa. Our results are compared to \textit{(i)} equivalent experimental results previously obtained on single layer \Ca\ and \textit{(ii)} local cluster modelling of \biCa\ and \Ca. In particular, the oxygen $K$-edge RIXS spectra are fundamentally different in \biCa\ and \Ca, reflecting their different ground states. Whereas in \Ca\ a set of excitations within the $t_{2g}$ subspace, consisting of two low-energy and two mid/high-energy structures, is observed, only the two lowest intra-$t_{2g}$ excitations have a significant amplitude in \biCa. This effect is captured by the local cluster modelling taking into account the different in-plane magnetic couplings. Finally, we demonstrated that the lowest lying intra-\tg\ excitation in \biCa\ is dispersing, revealing its collective origin. We argue based on the exact dispersion and comparison to spin correlation function computations, that this excitation is magnonic rather than phononic in nature. In fact, it is suggested to be dominantly a transverse mode with multi-particle character, which is indirectly allowed at the oxygen $K$-edge through substantial SOC of Ru ions.

\section{Appendix A: Model Hamiltonian} 
We report here the details of the microscopic model describing the energy levels and wave functions of the considered Ru-O-Ru cluster. The examined Hamiltonian~\cite{CuocoPRB06a,CuocoPRB06b} is expressed as:
\begin{eqnarray}
H=H_{kin}+H_{el-el}+H_{cf}+H_{soc}+H_{m} \,. \label{totham}
\end{eqnarray}
The first term in Eq. \ref{totham} is the kinetic operator describing the Ru-O connectivity:
\begin{eqnarray}
H_{kin}=\sum_{ij,\alpha \beta,\sigma} t^{\alpha \beta}_{ij}( p^{\dagger}_{{i} \alpha \sigma}
d_{{ j} \beta \sigma} +h.c.)\ ,
\end{eqnarray}
where $d^{\dagger}_{i \beta \sigma}$  is the creation operator for an
electron with spin $\sigma$ at the $i$ site in the $\beta$
orbital of the $t_{2g}$ sector ($d_{xy}$, $d_{xz}$, $d_{yz}$), while $p_{i \alpha \sigma}$ is the annihilation operator of an electron with spin $\sigma$ at the $i$ site in the $\alpha$
orbital of the ($p_{x}$, $p_{y}$, $p_{z}$) space of the oxygen. Hopping amplitudes $t^{\alpha \beta}_{ij}$ include all
the allowed symmetry terms according to the Slater-Koster rules~\cite{slaterPR1954,brzezickiPRL2015} for a given bond connecting a ruthenium to an
oxygen atom along, say, the $x$ direction.
We allow for the relative rotation of the oxygen octahedra surrounding the Ru site, assuming that the
Ru-O-Ru bond can form an angle $\theta$=(180\degree-$\phi$). The case
with $\phi$= 0\degree corresponds to the tetragonal undistorted bond,
while a non vanishing value of $\phi$ arises when the RuO$_6$ octahedra
are rotated of the corresponding angle around the $c$
axis.\par
The second term is the Coulomb interaction, which is expressed in terms of Kanamori parameters $U$, $U'$, and $J_{H}$ as follows
\begin{equation}
\begin{split}
H_{el-el} =  U \sum_{i\alpha} n_{{i} \alpha
\uparrow} n_{{i} \alpha \downarrow} - 2 J_{H} \sum_{i \alpha
\beta} {\bf S}_{{i} \alpha}\cdot {\bf S}_{{i} \beta}+\\
(U'-\frac{J_{H}}{2}) \sum_{i\alpha\neq\beta} n_{{i} \alpha}
n_{{i} \beta}+J' \sum_{i\alpha \beta} d^{\dagger}_{{i}
\alpha \uparrow} d^{\dagger}_{{i} \alpha \downarrow} d_{{\bf
i} \beta \uparrow} d_{{i} \beta \downarrow}
\end{split}
\end{equation}
\noindent where $n_{{i} \alpha \sigma}$, ${\bf S}_{{i}
\alpha}$ are the on site charge for spin $\sigma$ and the spin
operators for the $\alpha$ orbital, respectively. $U$ ($U'$) is
the intra (inter)- orbital Coulomb repulsion, $J_{H}$ is the Hund
coupling, and $J'$ the pair hopping term. Due to the invariance
for rotations in the orbital space, the following relations hold:
$U=U'+2 J_{H}$, $J'=J_{H}$.\par
The $H_{cf}$ part of the Hamiltonian $H$ is the crystalline field potential,
controlling the symmetry lowering from cubic to tetragonal one, due to the compression of RuO$_6$ octahedra along the $c$ axis.:
\begin{eqnarray}
H_{cf}= \sum_i \Delta_i [n_{i xy}-\frac{1}{2}(n_{{i}xz}+n_{ i yz})]
\,
\end{eqnarray}
The SOC Hamiltonian reads as
\begin{equation}
H_{soc}=\lambda \sum_i {\bf L}_i \cdot {\bf S}_i \, .
\end{equation}
Due to the cubic CF terms in RuO$_6$ octahedra separating the lower $t_{2g}$ from the unoccupied $e_g$ levels, $\bf L_{i}$ stands for the angular momentum operator projected onto the $t_{2g}$ subspace. Its components have the following expression in terms of orbital fermionic operators:
\begin{equation}
    \begin{split}
        L_{i x}=i \sum_{\sigma} [d^{\dagger}_{{ i}xy\sigma}
        d_{{ i}xz\sigma}- d^{\dagger}_{{ i}xz\sigma} d_{{i}xy\sigma}]\\
        L_{i y}=i \sum_{\sigma} [d^{\dagger}_{i yz \sigma}
        d_{ i xy \sigma}-d^{\dagger}_{i xy \sigma} d_{ i yz\sigma}]\\
        L_{i z}=i \sum_{\sigma} [d^{\dagger}_{ i xz\sigma}
        d_{ i yz\sigma}- d^{\dagger}_{ i yz\sigma} d_{ i xz\sigma}]
    \end{split}
\end{equation}
Finally, $H_{m}$ in Eq. \ref{totham} is a an effective exchange field which pins the magnetization at the Ru sites to be in the ($x,y$) plane for the FM ground state:
\begin{equation}
H_{m}=\sum_{i} \textbf{S}_i \cdot \textbf{B}_{xy} \,.
\end{equation}

\section{APPENDIX B: Calculation of the RIXS cross section}
The RIXS intensity is described by the
Kramers-Heisenberg relation \cite{} 
\begin{equation}
I(\omega,{\bf q},\epsilon,\epsilon^{'})=\sum_{f} |A_{fg}   (\omega,{\bf{q}},\epsilon,\epsilon^{'})|^{2} \delta(E_f+\omega_{k^{'}}-E_{g}-\omega_{k})
\end{equation}
where $\omega=\omega_{k^{'}}-\omega_{k}$ and and $\bf{q}=\bf{k^{'}}-\bf{k}$ stand for the energy and momentum transferred by the scattered photon, and $\epsilon$ and $\epsilon^{'}$ for the incoming and outgoing light polarization vectors.
We adopt the dipole and fast collision approximation\cite{}, in which the RIXS scattering amplitude  is reduced to 
\begin{equation}
A_{fg} =\frac{1}{i \Gamma} \langle f| R(\epsilon,\epsilon^{'},{\bf q}) |g\rangle\ ,
 \label{eq:amplitude}
\end{equation}
where $R$  is the effective RIXS scattering operator describing two subsequent dipole transitions, and $\Gamma$ is the core-hole broadening. In the oxygen $K$-edge RIXS, the dipole transitions create an O 1$s$ core hole and extra valence electron in a 2$p$ obital and viceversa, and the scattering operator has the following expression:
\begin{equation}
R(\epsilon_{\nu},\epsilon_{\nu^{'}}) \propto \sum_{i,\sigma} e^{i {\bf q}  \cdot {\bf r}_i} p_{\nu^{'} \sigma}p_{\nu \sigma}  ,
\end{equation}
where $\nu$ is the $(x, y, z) $ orbital and the sum over the different spin states is assumed. Matrix elements are then evaluated among oxygen valence states in Eq. \ref{eq:amplitude}. Notably, the valence electron in a 2$p$ obital hybridizes and interacts with the Ru $d$ electrons.\par
In the adopted experimental scattering geometry, the dependence upon the incident angle $\theta_{in}$ and scattering angle $\alpha=130\degree$ between the incoming/outgoing polariization vectors is:\\
\begin{equation}
\begin{split}
\epsilon_{LH}&=\epsilon_x \cos \theta_{in}+\epsilon_z \sin \theta_{in}\\
\epsilon_{LV}&=\epsilon_y\\    
\epsilon^{'}&=\epsilon^{'}_x \cos (\theta_{in}+\alpha)+\epsilon^{'}_y+\epsilon^{'}_z \sin(\theta_{in}+\alpha) 
\end{split}
\end{equation}
Here the coordinate frame $(x,y,z),$ corresponds to the tetragonal axis frame $(a_{T},b_{T},c)$. Since the outgoing polarization is not resolved, the RIXS intensity is obtained by summing up incoherently over all the three polarization directions $(\epsilon^{'}_x ,\epsilon^{'}_y,\epsilon^{'}_z)$.\par

\begin{figure}
	\begin{center}
		\includegraphics[width=0.45\textwidth]{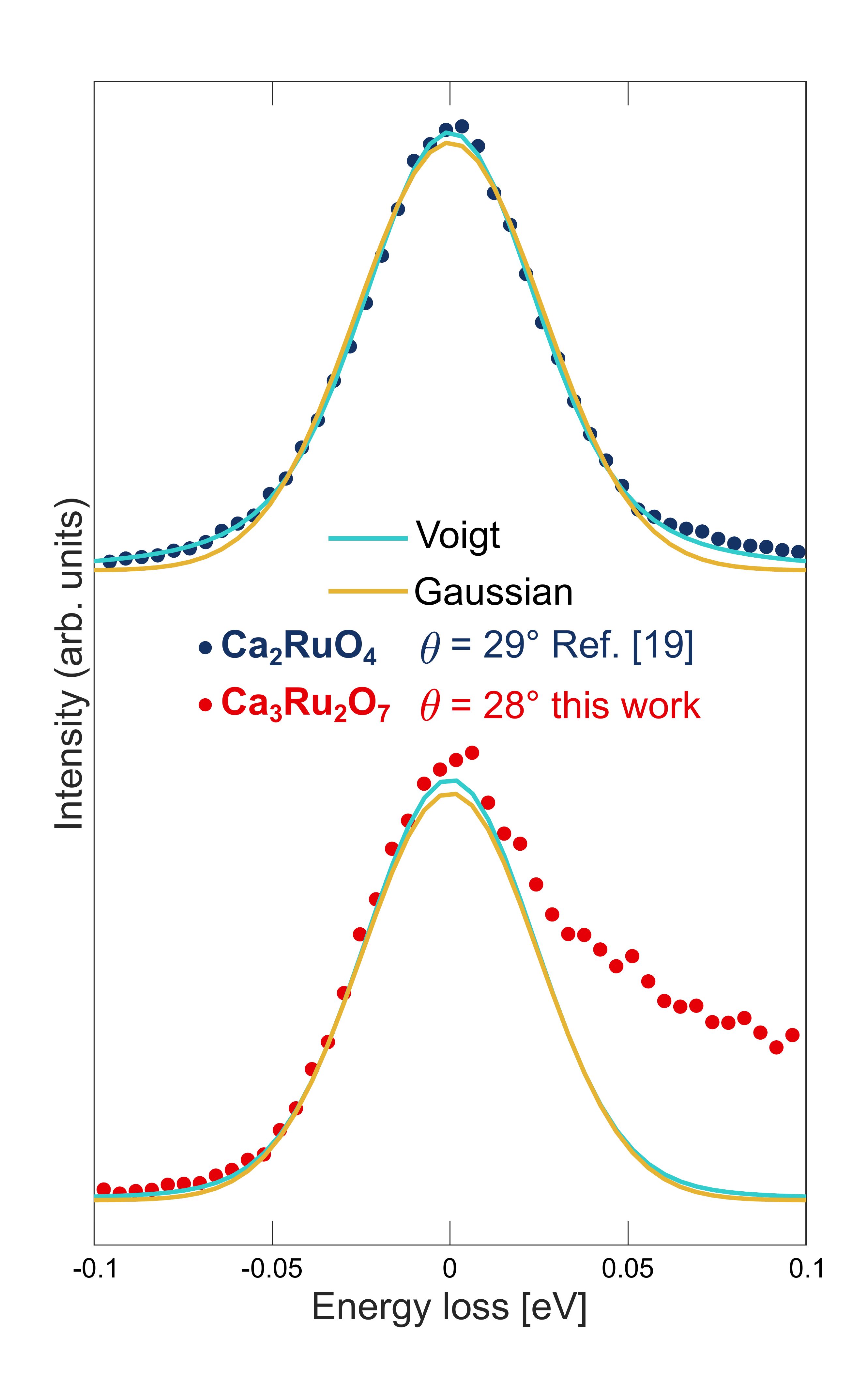}
	\end{center}
	\caption{Analysis of elastic scattering. Elastic scattering part of the RIXS spectra recorded on \Ca\ (top curve, taken from Ref.~\onlinecite{das_spin-orbital_2018}) and \biCa\ (bottom curve) near specular condition as indicated. Solid lines are Gaussian and Voigt fits. In both profiles, the Gaussian width is fixed to the energy resolution of the experiment. On \biCa, the full widths at half maximum (FWHM) of the Gaussian and Lorentzian contributions in the Voigt profile are 53.3~meV and 6.9~meV, respectively. To enhance visibility, the spectra are normalized to their maximum and the spectra and fits of \Ca\ are given a vertical shift.
	}
	\label{fig:fig6}
\end{figure}

\section{APPENDIX C: Subtraction of elastic scattering contributions}
The analysis of the lowest lying (dispersive)  A excitation in \biCa\ involves modelling of the elastic scattering. By assuming that the intrinsic elastic scattering line shape is Lorentzian and that the finite energy resolution is accounted by a Gaussian convolution, the effective line shape would therefore 
theoretically be a Voigt function. In practice, the Gaussian component is dominating over the Lorentzian. Therefore, both Gaussian and Voigt functions can be used to model elastic scattering.\par
In Fig.~\ref{fig:fig6}, elastic scattering profiles measured on \Ca\ and \biCa\ are shown and compared to fits using Gaussian and Voigt profiles. The counting statistic is better for the data recorded on \Ca. There, the fitting indeed suggests that the Voigt profile provides a slightly better description of the elastic line. This is to be expected as the fitting to a Voigt function involves and additional fitting parameter. For \biCa, the analysis is harder since (a) the counting statistics are less good and (b) the lowest lying excitation overlaps partially with the elastic scattering and (c) the elastic contribution in the spectra is overall much smaller than in \Ca. As shown in Fig.~\ref{fig:fig6}, both Voigt and Gaussian profiles provide 
good description of the elastic scattering. The difference between these two profiles are marginal and negligible compared to the intensity of the lowest lying excitation. To be consistent over all incident angles, the analysis of \biCa\ data presented in this paper made use of a Gaussian profile. Data presented in Ref.~\onlinecite{das_spin-orbital_2018} on \Ca\ is fitted to a Voigt function. The choice of Gaussian or Voigt function for the modelling of elastic scattering has no impact on the derived conclusions.

\section{Acknowlegdements}
The experimental work was performed at the ADRESS beamline of the SLS at the Paul Scherrer Institut, Villigen PSI, Switzerland. We thank the ADRESS beamline staff for technical support. K.v.A., M.H., Q.W, L.D., O.I., J.C. acknowledge support by the Swiss National Science Foundation. M.C. and F.F. acknowledge support by the project "Two-dimensional Oxides
Platform for SPIN-orbitronics nanotechnology (TOPSPIN)" funded by the MIUR (PRIN) Bando 2017 - Grant 20177SL7HC. Y.S. is funded by the Swedish Research Council (VR) with a Starting Grant (No.2017-05078). Work at the Paul Scherrer Institut has been funded by the Swiss National Science Foundation through the Sinergia network Mott Physics Beyond the Heisenberg (MPBH) model (SNSF Research grant numbers CRSII2\_141962 and CRSII2\_160765).

\bibliography{RIXS-Ca327}

\begin{thebibliography}{54}%
\makeatletter
\providecommand \@ifxundefined [1]{%
 \@ifx{#1\undefined}
}%
\providecommand \@ifnum [1]{%
 \ifnum #1\expandafter \@firstoftwo
 \else \expandafter \@secondoftwo
 \fi
}%
\providecommand \@ifx [1]{%
 \ifx #1\expandafter \@firstoftwo
 \else \expandafter \@secondoftwo
 \fi
}%
\providecommand \natexlab [1]{#1}%
\providecommand \enquote  [1]{``#1''}%
\providecommand \bibnamefont  [1]{#1}%
\providecommand \bibfnamefont [1]{#1}%
\providecommand \citenamefont [1]{#1}%
\providecommand \href@noop [0]{\@secondoftwo}%
\providecommand \href [0]{\begingroup \@sanitize@url \@href}%
\providecommand \@href[1]{\@@startlink{#1}\@@href}%
\providecommand \@@href[1]{\endgroup#1\@@endlink}%
\providecommand \@sanitize@url [0]{\catcode `\\12\catcode `\$12\catcode
  `\&12\catcode `\#12\catcode `\^12\catcode `\_12\catcode `\%12\relax}%
\providecommand \@@startlink[1]{}%
\providecommand \@@endlink[0]{}%
\providecommand \url  [0]{\begingroup\@sanitize@url \@url }%
\providecommand \@url [1]{\endgroup\@href {#1}{\urlprefix }}%
\providecommand \urlprefix  [0]{URL }%
\providecommand \Eprint [0]{\href }%
\providecommand \doibase [0]{http://dx.doi.org/}%
\providecommand \selectlanguage [0]{\@gobble}%
\providecommand \bibinfo  [0]{\@secondoftwo}%
\providecommand \bibfield  [0]{\@secondoftwo}%
\providecommand \translation [1]{[#1]}%
\providecommand \BibitemOpen [0]{}%
\providecommand \bibitemStop [0]{}%
\providecommand \bibitemNoStop [0]{.\EOS\space}%
\providecommand \EOS [0]{\spacefactor3000\relax}%
\providecommand \BibitemShut  [1]{\csname bibitem#1\endcsname}%
\let\auto@bib@innerbib\@empty
\bibitem [{\citenamefont {Karpus}\ \emph {et~al.}(2004)\citenamefont {Karpus},
  \citenamefont {Gupta}, \citenamefont {Barath}, \citenamefont {Cooper},\ and\
  \citenamefont {Cao}}]{KarpusPRL2004}%
  \BibitemOpen
  \bibfield  {author} {\bibinfo {author} {\bibfnamefont {J.~F.}\ \bibnamefont
  {Karpus}}, \bibinfo {author} {\bibfnamefont {R.}~\bibnamefont {Gupta}},
  \bibinfo {author} {\bibfnamefont {H.}~\bibnamefont {Barath}}, \bibinfo
  {author} {\bibfnamefont {S.~L.}\ \bibnamefont {Cooper}}, \ and\ \bibinfo
  {author} {\bibfnamefont {G.}~\bibnamefont {Cao}},\ }\href {\doibase
  10.1103/PhysRevLett.93.167205} {\bibfield  {journal} {\bibinfo  {journal}
  {Phys. Rev. Lett.}\ }\textbf {\bibinfo {volume} {93}},\ \bibinfo {pages}
  {167205} (\bibinfo {year} {2004})}\BibitemShut {NoStop}%
\bibitem [{\citenamefont {Lin}\ \emph {et~al.}(2005)\citenamefont {Lin},
  \citenamefont {Zhou}, \citenamefont {Durairaj}, \citenamefont {Schlottmann},\
  and\ \citenamefont {Cao}}]{LinPRL2005}%
  \BibitemOpen
  \bibfield  {author} {\bibinfo {author} {\bibfnamefont {X.~N.}\ \bibnamefont
  {Lin}}, \bibinfo {author} {\bibfnamefont {Z.~X.}\ \bibnamefont {Zhou}},
  \bibinfo {author} {\bibfnamefont {V.}~\bibnamefont {Durairaj}}, \bibinfo
  {author} {\bibfnamefont {P.}~\bibnamefont {Schlottmann}}, \ and\ \bibinfo
  {author} {\bibfnamefont {G.}~\bibnamefont {Cao}},\ }\href {\doibase
  10.1103/PhysRevLett.95.017203} {\bibfield  {journal} {\bibinfo  {journal}
  {Phys. Rev. Lett.}\ }\textbf {\bibinfo {volume} {95}},\ \bibinfo {pages}
  {017203} (\bibinfo {year} {2005})}\BibitemShut {NoStop}%
\bibitem [{\citenamefont {Bao}\ \emph {et~al.}(2008)\citenamefont {Bao},
  \citenamefont {Mao}, \citenamefont {Qu},\ and\ \citenamefont
  {Lynn}}]{BaoPRL2008}%
  \BibitemOpen
  \bibfield  {author} {\bibinfo {author} {\bibfnamefont {W.}~\bibnamefont
  {Bao}}, \bibinfo {author} {\bibfnamefont {Z.~Q.}\ \bibnamefont {Mao}},
  \bibinfo {author} {\bibfnamefont {Z.}~\bibnamefont {Qu}}, \ and\ \bibinfo
  {author} {\bibfnamefont {J.~W.}\ \bibnamefont {Lynn}},\ }\href {\doibase
  10.1103/PhysRevLett.100.247203} {\bibfield  {journal} {\bibinfo  {journal}
  {Phys. Rev. Lett.}\ }\textbf {\bibinfo {volume} {100}},\ \bibinfo {pages}
  {247203} (\bibinfo {year} {2008})}\BibitemShut {NoStop}%
\bibitem [{\citenamefont {Kikugawa}\ \emph {et~al.}(2010)\citenamefont
  {Kikugawa}, \citenamefont {Winfried~Rost}, \citenamefont {William~Hicks},
  \citenamefont {John~Schofield},\ and\ \citenamefont
  {Peter~Mackenzie}}]{KikugawaJPSJ2010}%
  \BibitemOpen
  \bibfield  {author} {\bibinfo {author} {\bibfnamefont {N.}~\bibnamefont
  {Kikugawa}}, \bibinfo {author} {\bibfnamefont {A.}~\bibnamefont
  {Winfried~Rost}}, \bibinfo {author} {\bibfnamefont {C.}~\bibnamefont
  {William~Hicks}}, \bibinfo {author} {\bibfnamefont {A.}~\bibnamefont
  {John~Schofield}}, \ and\ \bibinfo {author} {\bibfnamefont {A.}~\bibnamefont
  {Peter~Mackenzie}},\ }\href {\doibase 10.1143/JPSJ.79.024704} {\bibfield
  {journal} {\bibinfo  {journal} {J. Phys. Soc. Jpn.}\ }\textbf {\bibinfo
  {volume} {79}},\ \bibinfo {pages} {024704} (\bibinfo {year}
  {2010})}\BibitemShut {NoStop}%
\bibitem [{\citenamefont {Zhu}\ \emph {et~al.}(2016)\citenamefont {Zhu},
  \citenamefont {Peng}, \citenamefont {Zou}, \citenamefont {Prokes},
  \citenamefont {Mahanti}, \citenamefont {Hong}, \citenamefont {Mao},
  \citenamefont {Liu},\ and\ \citenamefont {Ke}}]{ZhuPRL2016}%
  \BibitemOpen
  \bibfield  {author} {\bibinfo {author} {\bibfnamefont {M.}~\bibnamefont
  {Zhu}}, \bibinfo {author} {\bibfnamefont {J.}~\bibnamefont {Peng}}, \bibinfo
  {author} {\bibfnamefont {T.}~\bibnamefont {Zou}}, \bibinfo {author}
  {\bibfnamefont {K.}~\bibnamefont {Prokes}}, \bibinfo {author} {\bibfnamefont
  {S.~D.}\ \bibnamefont {Mahanti}}, \bibinfo {author} {\bibfnamefont
  {T.}~\bibnamefont {Hong}}, \bibinfo {author} {\bibfnamefont {Z.~Q.}\
  \bibnamefont {Mao}}, \bibinfo {author} {\bibfnamefont {G.~Q.}\ \bibnamefont
  {Liu}}, \ and\ \bibinfo {author} {\bibfnamefont {X.}~\bibnamefont {Ke}},\
  }\href {\doibase 10.1103/PhysRevLett.116.216401} {\bibfield  {journal}
  {\bibinfo  {journal} {Phys. Rev. Lett.}\ }\textbf {\bibinfo {volume} {116}},\
  \bibinfo {pages} {216401} (\bibinfo {year} {2016})}\BibitemShut {NoStop}%
\bibitem [{\citenamefont {Xing}\ \emph {et~al.}(2018)\citenamefont {Xing},
  \citenamefont {Wen}, \citenamefont {Shen}, \citenamefont {He}, \citenamefont
  {Cai}, \citenamefont {Peng}, \citenamefont {Wang}, \citenamefont {Tian},
  \citenamefont {Xu}, \citenamefont {Ku}, \citenamefont {Mao},\ and\
  \citenamefont {Liu}}]{XingPRB2018}%
  \BibitemOpen
  \bibfield  {author} {\bibinfo {author} {\bibfnamefont {H.}~\bibnamefont
  {Xing}}, \bibinfo {author} {\bibfnamefont {L.}~\bibnamefont {Wen}}, \bibinfo
  {author} {\bibfnamefont {C.}~\bibnamefont {Shen}}, \bibinfo {author}
  {\bibfnamefont {J.}~\bibnamefont {He}}, \bibinfo {author} {\bibfnamefont
  {X.}~\bibnamefont {Cai}}, \bibinfo {author} {\bibfnamefont {J.}~\bibnamefont
  {Peng}}, \bibinfo {author} {\bibfnamefont {S.}~\bibnamefont {Wang}}, \bibinfo
  {author} {\bibfnamefont {M.}~\bibnamefont {Tian}}, \bibinfo {author}
  {\bibfnamefont {Z.-A.}\ \bibnamefont {Xu}}, \bibinfo {author} {\bibfnamefont
  {W.}~\bibnamefont {Ku}}, \bibinfo {author} {\bibfnamefont {Z.}~\bibnamefont
  {Mao}}, \ and\ \bibinfo {author} {\bibfnamefont {Y.}~\bibnamefont {Liu}},\
  }\href {\doibase 10.1103/PhysRevB.97.041113} {\bibfield  {journal} {\bibinfo
  {journal} {Phys. Rev. B}\ }\textbf {\bibinfo {volume} {97}},\ \bibinfo
  {pages} {041113} (\bibinfo {year} {2018})}\BibitemShut {NoStop}%
\bibitem [{\citenamefont {Sow}\ \emph {et~al.}(2019)\citenamefont {Sow},
  \citenamefont {Numasaki}, \citenamefont {Mattoni}, \citenamefont {Yonezawa},
  \citenamefont {Kikugawa}, \citenamefont {Uji},\ and\ \citenamefont
  {Maeno}}]{SowPRL2019}%
  \BibitemOpen
  \bibfield  {author} {\bibinfo {author} {\bibfnamefont {C.}~\bibnamefont
  {Sow}}, \bibinfo {author} {\bibfnamefont {R.}~\bibnamefont {Numasaki}},
  \bibinfo {author} {\bibfnamefont {G.}~\bibnamefont {Mattoni}}, \bibinfo
  {author} {\bibfnamefont {S.}~\bibnamefont {Yonezawa}}, \bibinfo {author}
  {\bibfnamefont {N.}~\bibnamefont {Kikugawa}}, \bibinfo {author}
  {\bibfnamefont {S.}~\bibnamefont {Uji}}, \ and\ \bibinfo {author}
  {\bibfnamefont {Y.}~\bibnamefont {Maeno}},\ }\href {\doibase
  10.1103/PhysRevLett.122.196602} {\bibfield  {journal} {\bibinfo  {journal}
  {Phys. Rev. Lett.}\ }\textbf {\bibinfo {volume} {122}},\ \bibinfo {pages}
  {196602} (\bibinfo {year} {2019})}\BibitemShut {NoStop}%
\bibitem [{\citenamefont {Puggioni}\ \emph {et~al.}(2020)\citenamefont
  {Puggioni}, \citenamefont {Horio}, \citenamefont {Chang},\ and\ \citenamefont
  {Rondinelli}}]{PuggioniPRR2020}%
  \BibitemOpen
  \bibfield  {author} {\bibinfo {author} {\bibfnamefont {D.}~\bibnamefont
  {Puggioni}}, \bibinfo {author} {\bibfnamefont {M.}~\bibnamefont {Horio}},
  \bibinfo {author} {\bibfnamefont {J.}~\bibnamefont {Chang}}, \ and\ \bibinfo
  {author} {\bibfnamefont {J.~M.}\ \bibnamefont {Rondinelli}},\ }\href
  {\doibase 10.1103/PhysRevResearch.2.023141} {\bibfield  {journal} {\bibinfo
  {journal} {Phys. Rev. Research}\ }\textbf {\bibinfo {volume} {2}},\ \bibinfo
  {pages} {023141} (\bibinfo {year} {2020})}\BibitemShut {NoStop}%
\bibitem [{\citenamefont {Cao}\ \emph {et~al.}(2003)\citenamefont {Cao},
  \citenamefont {Balicas}, \citenamefont {Xin}, \citenamefont {Crow},\ and\
  \citenamefont {Nelson}}]{cao_quantum_2003}%
  \BibitemOpen
  \bibfield  {author} {\bibinfo {author} {\bibfnamefont {G.}~\bibnamefont
  {Cao}}, \bibinfo {author} {\bibfnamefont {L.}~\bibnamefont {Balicas}},
  \bibinfo {author} {\bibfnamefont {Y.}~\bibnamefont {Xin}}, \bibinfo {author}
  {\bibfnamefont {J.~E.}\ \bibnamefont {Crow}}, \ and\ \bibinfo {author}
  {\bibfnamefont {C.~S.}\ \bibnamefont {Nelson}},\ }\href {\doibase
  10.1103/PhysRevB.67.184405} {\bibfield  {journal} {\bibinfo  {journal} {Phys.
  Rev. B}\ }\textbf {\bibinfo {volume} {67}},\ \bibinfo {pages} {184405}
  (\bibinfo {year} {2003})}\BibitemShut {NoStop}%
\bibitem [{\citenamefont {Yoshida}\ \emph {et~al.}(2004)\citenamefont
  {Yoshida}, \citenamefont {Nagai}, \citenamefont {Ikeda}, \citenamefont
  {Shirakawa}, \citenamefont {Kosaka},\ and\ \citenamefont
  {Môri}}]{yoshida_quasi-two-dimensional_2004}%
  \BibitemOpen
  \bibfield  {author} {\bibinfo {author} {\bibfnamefont {Y.}~\bibnamefont
  {Yoshida}}, \bibinfo {author} {\bibfnamefont {I.}~\bibnamefont {Nagai}},
  \bibinfo {author} {\bibfnamefont {S.-I.}\ \bibnamefont {Ikeda}}, \bibinfo
  {author} {\bibfnamefont {N.}~\bibnamefont {Shirakawa}}, \bibinfo {author}
  {\bibfnamefont {M.}~\bibnamefont {Kosaka}}, \ and\ \bibinfo {author}
  {\bibfnamefont {N.}~\bibnamefont {Môri}},\ }\href {\doibase
  10.1103/PhysRevB.69.220411} {\bibfield  {journal} {\bibinfo  {journal} {Phys.
  Rev. B}\ }\textbf {\bibinfo {volume} {69}},\ \bibinfo {pages} {220411}
  (\bibinfo {year} {2004})}\BibitemShut {NoStop}%
\bibitem [{\citenamefont {Ohmichi}\ \emph {et~al.}(2004)\citenamefont
  {Ohmichi}, \citenamefont {Yoshida}, \citenamefont {Ikeda}, \citenamefont
  {Shirakawa},\ and\ \citenamefont {Osada}}]{ohmichi_colossal_2004}%
  \BibitemOpen
  \bibfield  {author} {\bibinfo {author} {\bibfnamefont {E.}~\bibnamefont
  {Ohmichi}}, \bibinfo {author} {\bibfnamefont {Y.}~\bibnamefont {Yoshida}},
  \bibinfo {author} {\bibfnamefont {S.~I.}\ \bibnamefont {Ikeda}}, \bibinfo
  {author} {\bibfnamefont {N.}~\bibnamefont {Shirakawa}}, \ and\ \bibinfo
  {author} {\bibfnamefont {T.}~\bibnamefont {Osada}},\ }\href {\doibase
  10.1103/PhysRevB.70.104414} {\bibfield  {journal} {\bibinfo  {journal} {Phys.
  Rev. B}\ }\textbf {\bibinfo {volume} {70}},\ \bibinfo {pages} {104414}
  (\bibinfo {year} {2004})}\BibitemShut {NoStop}%
\bibitem [{\citenamefont {Yoshida}\ \emph {et~al.}(2005)\citenamefont
  {Yoshida}, \citenamefont {Ikeda}, \citenamefont {Matsuhata}, \citenamefont
  {Shirakawa}, \citenamefont {Lee},\ and\ \citenamefont
  {Katano}}]{yoshidaPRB2005}%
  \BibitemOpen
  \bibfield  {author} {\bibinfo {author} {\bibfnamefont {Y.}~\bibnamefont
  {Yoshida}}, \bibinfo {author} {\bibfnamefont {S.-I.}\ \bibnamefont {Ikeda}},
  \bibinfo {author} {\bibfnamefont {H.}~\bibnamefont {Matsuhata}}, \bibinfo
  {author} {\bibfnamefont {N.}~\bibnamefont {Shirakawa}}, \bibinfo {author}
  {\bibfnamefont {C.~H.}\ \bibnamefont {Lee}}, \ and\ \bibinfo {author}
  {\bibfnamefont {S.}~\bibnamefont {Katano}},\ }\href {\doibase
  10.1103/PhysRevB.72.054412} {\bibfield  {journal} {\bibinfo  {journal} {Phys.
  Rev. B}\ }\textbf {\bibinfo {volume} {72}},\ \bibinfo {pages} {054412}
  (\bibinfo {year} {2005})}\BibitemShut {NoStop}%
\bibitem [{\citenamefont {Nakatsuji}\ \emph {et~al.}(2004)\citenamefont
  {Nakatsuji}, \citenamefont {Dobrosavljevi\ifmmode~\acute{c}\else \'{c}\fi{}},
  \citenamefont {Tanaskovi\ifmmode~\acute{c}\else \'{c}\fi{}}, \citenamefont
  {Minakata}, \citenamefont {Fukazawa},\ and\ \citenamefont
  {Maeno}}]{NakatsujiPRL2004}%
  \BibitemOpen
  \bibfield  {author} {\bibinfo {author} {\bibfnamefont {S.}~\bibnamefont
  {Nakatsuji}}, \bibinfo {author} {\bibfnamefont {V.}~\bibnamefont
  {Dobrosavljevi\ifmmode~\acute{c}\else \'{c}\fi{}}}, \bibinfo {author}
  {\bibfnamefont {D.}~\bibnamefont {Tanaskovi\ifmmode~\acute{c}\else
  \'{c}\fi{}}}, \bibinfo {author} {\bibfnamefont {M.}~\bibnamefont {Minakata}},
  \bibinfo {author} {\bibfnamefont {H.}~\bibnamefont {Fukazawa}}, \ and\
  \bibinfo {author} {\bibfnamefont {Y.}~\bibnamefont {Maeno}},\ }\href
  {\doibase 10.1103/PhysRevLett.93.146401} {\bibfield  {journal} {\bibinfo
  {journal} {Phys. Rev. Lett.}\ }\textbf {\bibinfo {volume} {93}},\ \bibinfo
  {pages} {146401} (\bibinfo {year} {2004})}\BibitemShut {NoStop}%
\bibitem [{\citenamefont {Takenaka}\ \emph {et~al.}(2017)\citenamefont
  {Takenaka}, \citenamefont {Okamoto}, \citenamefont {Shinoda}, \citenamefont
  {Katayama},\ and\ \citenamefont {Sakai}}]{TakenakaNatComm2017}%
  \BibitemOpen
  \bibfield  {author} {\bibinfo {author} {\bibfnamefont {K.}~\bibnamefont
  {Takenaka}}, \bibinfo {author} {\bibfnamefont {Y.}~\bibnamefont {Okamoto}},
  \bibinfo {author} {\bibfnamefont {T.}~\bibnamefont {Shinoda}}, \bibinfo
  {author} {\bibfnamefont {N.}~\bibnamefont {Katayama}}, \ and\ \bibinfo
  {author} {\bibfnamefont {Y.}~\bibnamefont {Sakai}},\ }\href {\doibase
  10.1038/ncomms14102} {\bibfield  {journal} {\bibinfo  {journal} {Nature
  Communications}\ }\textbf {\bibinfo {volume} {8}},\ \bibinfo {pages} {14102}
  (\bibinfo {year} {2017})}\BibitemShut {NoStop}%
\bibitem [{\citenamefont {Friedt}\ \emph {et~al.}(2001)\citenamefont {Friedt},
  \citenamefont {Braden}, \citenamefont {Andr\'{e}}, \citenamefont {Adelmann},
  \citenamefont {Nakatsuji},\ and\ \citenamefont
  {Maeno}}]{friedt_structural_2001}%
  \BibitemOpen
  \bibfield  {author} {\bibinfo {author} {\bibfnamefont {O.}~\bibnamefont
  {Friedt}}, \bibinfo {author} {\bibfnamefont {M.}~\bibnamefont {Braden}},
  \bibinfo {author} {\bibfnamefont {G.}~\bibnamefont {Andr\'{e}}}, \bibinfo
  {author} {\bibfnamefont {P.}~\bibnamefont {Adelmann}}, \bibinfo {author}
  {\bibfnamefont {S.}~\bibnamefont {Nakatsuji}}, \ and\ \bibinfo {author}
  {\bibfnamefont {Y.}~\bibnamefont {Maeno}},\ }\href {\doibase
  10.1103/PhysRevB.63.174432} {\bibfield  {journal} {\bibinfo  {journal} {Phys.
  Rev. B}\ }\textbf {\bibinfo {volume} {63}},\ \bibinfo {pages} {174432}
  (\bibinfo {year} {2001})}\BibitemShut {NoStop}%
\bibitem [{\citenamefont {Braden}\ \emph {et~al.}(1998)\citenamefont {Braden},
  \citenamefont {André}, \citenamefont {Nakatsuji},\ and\ \citenamefont
  {Maeno}}]{braden_crystal_1998}%
  \BibitemOpen
  \bibfield  {author} {\bibinfo {author} {\bibfnamefont {M.}~\bibnamefont
  {Braden}}, \bibinfo {author} {\bibfnamefont {G.}~\bibnamefont {André}},
  \bibinfo {author} {\bibfnamefont {S.}~\bibnamefont {Nakatsuji}}, \ and\
  \bibinfo {author} {\bibfnamefont {Y.}~\bibnamefont {Maeno}},\ }\href
  {\doibase 10.1103/PhysRevB.58.847} {\bibfield  {journal} {\bibinfo  {journal}
  {Phys. Rev. B}\ }\textbf {\bibinfo {volume} {58}},\ \bibinfo {pages} {847}
  (\bibinfo {year} {1998})}\BibitemShut {NoStop}%
\bibitem [{\citenamefont {Jain}\ \emph {et~al.}(2017)\citenamefont {Jain},
  \citenamefont {Krautloher}, \citenamefont {Porras}, \citenamefont {Ryu},
  \citenamefont {Chen}, \citenamefont {Abernathy}, \citenamefont {Park},
  \citenamefont {Ivanov}, \citenamefont {Chaloupka}, \citenamefont
  {Khaliullin}, \citenamefont {Keimer},\ and\ \citenamefont
  {Kim}}]{jain2017higgs}%
  \BibitemOpen
  \bibfield  {author} {\bibinfo {author} {\bibfnamefont {A.}~\bibnamefont
  {Jain}}, \bibinfo {author} {\bibfnamefont {M.}~\bibnamefont {Krautloher}},
  \bibinfo {author} {\bibfnamefont {J.}~\bibnamefont {Porras}}, \bibinfo
  {author} {\bibfnamefont {G.~H.}\ \bibnamefont {Ryu}}, \bibinfo {author}
  {\bibfnamefont {D.~P.}\ \bibnamefont {Chen}}, \bibinfo {author}
  {\bibfnamefont {D.~L.}\ \bibnamefont {Abernathy}}, \bibinfo {author}
  {\bibfnamefont {J.~T.}\ \bibnamefont {Park}}, \bibinfo {author}
  {\bibfnamefont {A.}~\bibnamefont {Ivanov}}, \bibinfo {author} {\bibfnamefont
  {J.}~\bibnamefont {Chaloupka}}, \bibinfo {author} {\bibfnamefont
  {G.}~\bibnamefont {Khaliullin}}, \bibinfo {author} {\bibfnamefont
  {B.}~\bibnamefont {Keimer}}, \ and\ \bibinfo {author} {\bibfnamefont {B.~J.}\
  \bibnamefont {Kim}},\ }\href {\doibase 10.1038/nphys4077} {\bibfield
  {journal} {\bibinfo  {journal} {Nat. Phys.}\ }\textbf {\bibinfo {volume}
  {13}},\ \bibinfo {pages} {633} (\bibinfo {year} {2017})}\BibitemShut
  {NoStop}%
\bibitem [{\citenamefont {Souliou}\ \emph {et~al.}(2017)\citenamefont
  {Souliou}, \citenamefont {Chaloupka}, \citenamefont {Khaliullin},
  \citenamefont {Ryu}, \citenamefont {Jain}, \citenamefont {Kim}, \citenamefont
  {Le~Tacon},\ and\ \citenamefont {Keimer}}]{souliou_raman_2017}%
  \BibitemOpen
  \bibfield  {author} {\bibinfo {author} {\bibfnamefont {S.-M.}\ \bibnamefont
  {Souliou}}, \bibinfo {author} {\bibfnamefont {J.}~\bibnamefont {Chaloupka}},
  \bibinfo {author} {\bibfnamefont {G.}~\bibnamefont {Khaliullin}}, \bibinfo
  {author} {\bibfnamefont {G.}~\bibnamefont {Ryu}}, \bibinfo {author}
  {\bibfnamefont {A.}~\bibnamefont {Jain}}, \bibinfo {author} {\bibfnamefont
  {B.~J.}\ \bibnamefont {Kim}}, \bibinfo {author} {\bibfnamefont
  {M.}~\bibnamefont {Le~Tacon}}, \ and\ \bibinfo {author} {\bibfnamefont
  {B.}~\bibnamefont {Keimer}},\ }\href {\doibase
  10.1103/PhysRevLett.119.067201} {\bibfield  {journal} {\bibinfo  {journal}
  {Phys. Rev. Lett.}\ }\textbf {\bibinfo {volume} {119}},\ \bibinfo {pages}
  {067201} (\bibinfo {year} {2017})}\BibitemShut {NoStop}%
\bibitem [{\citenamefont {Das}\ \emph {et~al.}(2018)\citenamefont {Das},
  \citenamefont {Forte}, \citenamefont {Fittipaldi}, \citenamefont {Fatuzzo},
  \citenamefont {Granata}, \citenamefont {Ivashko}, \citenamefont {Horio},
  \citenamefont {Schindler}, \citenamefont {Dantz}, \citenamefont {Tseng},
  \citenamefont {McNally}, \citenamefont {R\o{}nnow}, \citenamefont {Wan},
  \citenamefont {Christensen}, \citenamefont {Pelliciari}, \citenamefont
  {Olalde-Velasco}, \citenamefont {Kikugawa}, \citenamefont {Neupert},
  \citenamefont {Vecchione}, \citenamefont {Schmitt}, \citenamefont {Cuoco},\
  and\ \citenamefont {Chang}}]{das_spin-orbital_2018}%
  \BibitemOpen
  \bibfield  {author} {\bibinfo {author} {\bibfnamefont {L.}~\bibnamefont
  {Das}}, \bibinfo {author} {\bibfnamefont {F.}~\bibnamefont {Forte}}, \bibinfo
  {author} {\bibfnamefont {R.}~\bibnamefont {Fittipaldi}}, \bibinfo {author}
  {\bibfnamefont {C.~G.}\ \bibnamefont {Fatuzzo}}, \bibinfo {author}
  {\bibfnamefont {V.}~\bibnamefont {Granata}}, \bibinfo {author} {\bibfnamefont
  {O.}~\bibnamefont {Ivashko}}, \bibinfo {author} {\bibfnamefont
  {M.}~\bibnamefont {Horio}}, \bibinfo {author} {\bibfnamefont
  {F.}~\bibnamefont {Schindler}}, \bibinfo {author} {\bibfnamefont
  {M.}~\bibnamefont {Dantz}}, \bibinfo {author} {\bibfnamefont
  {Y.}~\bibnamefont {Tseng}}, \bibinfo {author} {\bibfnamefont {D.~E.}\
  \bibnamefont {McNally}}, \bibinfo {author} {\bibfnamefont {H.~M.}\
  \bibnamefont {R\o{}nnow}}, \bibinfo {author} {\bibfnamefont {W.}~\bibnamefont
  {Wan}}, \bibinfo {author} {\bibfnamefont {N.~B.}\ \bibnamefont
  {Christensen}}, \bibinfo {author} {\bibfnamefont {J.}~\bibnamefont
  {Pelliciari}}, \bibinfo {author} {\bibfnamefont {P.}~\bibnamefont
  {Olalde-Velasco}}, \bibinfo {author} {\bibfnamefont {N.}~\bibnamefont
  {Kikugawa}}, \bibinfo {author} {\bibfnamefont {T.}~\bibnamefont {Neupert}},
  \bibinfo {author} {\bibfnamefont {A.}~\bibnamefont {Vecchione}}, \bibinfo
  {author} {\bibfnamefont {T.}~\bibnamefont {Schmitt}}, \bibinfo {author}
  {\bibfnamefont {M.}~\bibnamefont {Cuoco}}, \ and\ \bibinfo {author}
  {\bibfnamefont {J.}~\bibnamefont {Chang}},\ }\href {\doibase
  10.1103/PhysRevX.8.011048} {\bibfield  {journal} {\bibinfo  {journal} {Phys.
  Rev. X}\ }\textbf {\bibinfo {volume} {8}},\ \bibinfo {pages} {011048}
  (\bibinfo {year} {2018})}\BibitemShut {NoStop}%
\bibitem [{\citenamefont {Fatuzzo}\ \emph {et~al.}(2015)\citenamefont
  {Fatuzzo}, \citenamefont {Dantz}, \citenamefont {Fatale}, \citenamefont
  {Olalde-Velasco}, \citenamefont {Shaik}, \citenamefont {Dalla~Piazza},
  \citenamefont {Toth}, \citenamefont {Pelliciari}, \citenamefont {Fittipaldi},
  \citenamefont {Vecchione}, \citenamefont {Kikugawa}, \citenamefont {Brooks},
  \citenamefont {R\o{}nnow}, \citenamefont {Grioni}, \citenamefont {R\"uegg},
  \citenamefont {Schmitt},\ and\ \citenamefont {Chang}}]{FatuzzoPRB2015}%
  \BibitemOpen
  \bibfield  {author} {\bibinfo {author} {\bibfnamefont {C.~G.}\ \bibnamefont
  {Fatuzzo}}, \bibinfo {author} {\bibfnamefont {M.}~\bibnamefont {Dantz}},
  \bibinfo {author} {\bibfnamefont {S.}~\bibnamefont {Fatale}}, \bibinfo
  {author} {\bibfnamefont {P.}~\bibnamefont {Olalde-Velasco}}, \bibinfo
  {author} {\bibfnamefont {N.~E.}\ \bibnamefont {Shaik}}, \bibinfo {author}
  {\bibfnamefont {B.}~\bibnamefont {Dalla~Piazza}}, \bibinfo {author}
  {\bibfnamefont {S.}~\bibnamefont {Toth}}, \bibinfo {author} {\bibfnamefont
  {J.}~\bibnamefont {Pelliciari}}, \bibinfo {author} {\bibfnamefont
  {R.}~\bibnamefont {Fittipaldi}}, \bibinfo {author} {\bibfnamefont
  {A.}~\bibnamefont {Vecchione}}, \bibinfo {author} {\bibfnamefont
  {N.}~\bibnamefont {Kikugawa}}, \bibinfo {author} {\bibfnamefont {J.~S.}\
  \bibnamefont {Brooks}}, \bibinfo {author} {\bibfnamefont {H.~M.}\
  \bibnamefont {R\o{}nnow}}, \bibinfo {author} {\bibfnamefont {M.}~\bibnamefont
  {Grioni}}, \bibinfo {author} {\bibfnamefont {C.}~\bibnamefont {R\"uegg}},
  \bibinfo {author} {\bibfnamefont {T.}~\bibnamefont {Schmitt}}, \ and\
  \bibinfo {author} {\bibfnamefont {J.}~\bibnamefont {Chang}},\ }\href
  {\doibase 10.1103/PhysRevB.91.155104} {\bibfield  {journal} {\bibinfo
  {journal} {Phys. Rev. B}\ }\textbf {\bibinfo {volume} {91}},\ \bibinfo
  {pages} {155104} (\bibinfo {year} {2015})}\BibitemShut {NoStop}%
\bibitem [{\citenamefont {Bisogni}\ \emph {et~al.}(2012)\citenamefont
  {Bisogni}, \citenamefont {Simonelli}, \citenamefont {Ament}, \citenamefont
  {Forte}, \citenamefont {Moretti~Sala}, \citenamefont {Minola}, \citenamefont
  {Huotari}, \citenamefont {van~den Brink}, \citenamefont {Ghiringhelli},
  \citenamefont {Brookes},\ and\ \citenamefont
  {Braicovich}}]{bisogni_bimagnon_2012}%
  \BibitemOpen
  \bibfield  {author} {\bibinfo {author} {\bibfnamefont {V.}~\bibnamefont
  {Bisogni}}, \bibinfo {author} {\bibfnamefont {L.}~\bibnamefont {Simonelli}},
  \bibinfo {author} {\bibfnamefont {L.~J.~P.}\ \bibnamefont {Ament}}, \bibinfo
  {author} {\bibfnamefont {F.}~\bibnamefont {Forte}}, \bibinfo {author}
  {\bibfnamefont {M.}~\bibnamefont {Moretti~Sala}}, \bibinfo {author}
  {\bibfnamefont {M.}~\bibnamefont {Minola}}, \bibinfo {author} {\bibfnamefont
  {S.}~\bibnamefont {Huotari}}, \bibinfo {author} {\bibfnamefont
  {J.}~\bibnamefont {van~den Brink}}, \bibinfo {author} {\bibfnamefont
  {G.}~\bibnamefont {Ghiringhelli}}, \bibinfo {author} {\bibfnamefont {N.~B.}\
  \bibnamefont {Brookes}}, \ and\ \bibinfo {author} {\bibfnamefont
  {L.}~\bibnamefont {Braicovich}},\ }\href {\doibase
  10.1103/PhysRevB.85.214527} {\bibfield  {journal} {\bibinfo  {journal} {Phys.
  Rev. B}\ }\textbf {\bibinfo {volume} {85}},\ \bibinfo {pages} {214527}
  (\bibinfo {year} {2012})}\BibitemShut {NoStop}%
\bibitem [{\citenamefont {Moretti~Sala}\ \emph {et~al.}(2014)\citenamefont
  {Moretti~Sala}, \citenamefont {Rossi}, \citenamefont {Boseggia},
  \citenamefont {Akimitsu}, \citenamefont {Brookes}, \citenamefont {Isobe},
  \citenamefont {Minola}, \citenamefont {Okabe}, \citenamefont {R\o{}nnow},
  \citenamefont {Simonelli}, \citenamefont {McMorrow},\ and\ \citenamefont
  {Monaco}}]{salaPRB2014}%
  \BibitemOpen
  \bibfield  {author} {\bibinfo {author} {\bibfnamefont {M.}~\bibnamefont
  {Moretti~Sala}}, \bibinfo {author} {\bibfnamefont {M.}~\bibnamefont {Rossi}},
  \bibinfo {author} {\bibfnamefont {S.}~\bibnamefont {Boseggia}}, \bibinfo
  {author} {\bibfnamefont {J.}~\bibnamefont {Akimitsu}}, \bibinfo {author}
  {\bibfnamefont {N.~B.}\ \bibnamefont {Brookes}}, \bibinfo {author}
  {\bibfnamefont {M.}~\bibnamefont {Isobe}}, \bibinfo {author} {\bibfnamefont
  {M.}~\bibnamefont {Minola}}, \bibinfo {author} {\bibfnamefont
  {H.}~\bibnamefont {Okabe}}, \bibinfo {author} {\bibfnamefont {H.~M.}\
  \bibnamefont {R\o{}nnow}}, \bibinfo {author} {\bibfnamefont {L.}~\bibnamefont
  {Simonelli}}, \bibinfo {author} {\bibfnamefont {D.~F.}\ \bibnamefont
  {McMorrow}}, \ and\ \bibinfo {author} {\bibfnamefont {G.}~\bibnamefont
  {Monaco}},\ }\href {\doibase 10.1103/PhysRevB.89.121101} {\bibfield
  {journal} {\bibinfo  {journal} {Phys. Rev. B}\ }\textbf {\bibinfo {volume}
  {89}},\ \bibinfo {pages} {121101} (\bibinfo {year} {2014})}\BibitemShut
  {NoStop}%
\bibitem [{\citenamefont {Lu}\ \emph {et~al.}(2018)\citenamefont {Lu},
  \citenamefont {Olalde-Velasco}, \citenamefont {Huang}, \citenamefont
  {Bisogni}, \citenamefont {Pelliciari}, \citenamefont {Fatale}, \citenamefont
  {Dantz}, \citenamefont {Vale}, \citenamefont {Hunter}, \citenamefont {Chang},
  \citenamefont {Strocov}, \citenamefont {Perry}, \citenamefont {Grioni},
  \citenamefont {McMorrow}, \citenamefont {R\o{}nnow},\ and\ \citenamefont
  {Schmitt}}]{lu_dispersive_2018}%
  \BibitemOpen
  \bibfield  {author} {\bibinfo {author} {\bibfnamefont {X.}~\bibnamefont
  {Lu}}, \bibinfo {author} {\bibfnamefont {P.}~\bibnamefont {Olalde-Velasco}},
  \bibinfo {author} {\bibfnamefont {Y.}~\bibnamefont {Huang}}, \bibinfo
  {author} {\bibfnamefont {V.}~\bibnamefont {Bisogni}}, \bibinfo {author}
  {\bibfnamefont {J.}~\bibnamefont {Pelliciari}}, \bibinfo {author}
  {\bibfnamefont {S.}~\bibnamefont {Fatale}}, \bibinfo {author} {\bibfnamefont
  {M.}~\bibnamefont {Dantz}}, \bibinfo {author} {\bibfnamefont {J.~G.}\
  \bibnamefont {Vale}}, \bibinfo {author} {\bibfnamefont {E.~C.}\ \bibnamefont
  {Hunter}}, \bibinfo {author} {\bibfnamefont {J.}~\bibnamefont {Chang}},
  \bibinfo {author} {\bibfnamefont {V.~N.}\ \bibnamefont {Strocov}}, \bibinfo
  {author} {\bibfnamefont {R.~S.}\ \bibnamefont {Perry}}, \bibinfo {author}
  {\bibfnamefont {M.}~\bibnamefont {Grioni}}, \bibinfo {author} {\bibfnamefont
  {D.~F.}\ \bibnamefont {McMorrow}}, \bibinfo {author} {\bibfnamefont {H.~M.}\
  \bibnamefont {R\o{}nnow}}, \ and\ \bibinfo {author} {\bibfnamefont
  {T.}~\bibnamefont {Schmitt}},\ }\href {\doibase 10.1103/PhysRevB.97.041102}
  {\bibfield  {journal} {\bibinfo  {journal} {Phys. Rev. B}\ }\textbf {\bibinfo
  {volume} {97}},\ \bibinfo {pages} {041102} (\bibinfo {year}
  {2018})}\BibitemShut {NoStop}%
\bibitem [{\citenamefont {Pincini}\ \emph {et~al.}(2019)\citenamefont
  {Pincini}, \citenamefont {Veiga}, \citenamefont {Dashwood}, \citenamefont
  {Forte}, \citenamefont {Cuoco}, \citenamefont {Perry}, \citenamefont
  {Bencok}, \citenamefont {Boothroyd},\ and\ \citenamefont
  {McMorrow}}]{pinciniPRB2019}%
  \BibitemOpen
  \bibfield  {author} {\bibinfo {author} {\bibfnamefont {D.}~\bibnamefont
  {Pincini}}, \bibinfo {author} {\bibfnamefont {L.~S.~I.}\ \bibnamefont
  {Veiga}}, \bibinfo {author} {\bibfnamefont {C.~D.}\ \bibnamefont {Dashwood}},
  \bibinfo {author} {\bibfnamefont {F.}~\bibnamefont {Forte}}, \bibinfo
  {author} {\bibfnamefont {M.}~\bibnamefont {Cuoco}}, \bibinfo {author}
  {\bibfnamefont {R.~S.}\ \bibnamefont {Perry}}, \bibinfo {author}
  {\bibfnamefont {P.}~\bibnamefont {Bencok}}, \bibinfo {author} {\bibfnamefont
  {A.~T.}\ \bibnamefont {Boothroyd}}, \ and\ \bibinfo {author} {\bibfnamefont
  {D.~F.}\ \bibnamefont {McMorrow}},\ }\href {\doibase
  https://doi.org/10.1103/PhysRevB.99.075125} {\bibfield  {journal} {\bibinfo
  {journal} {Phys. Rev. B}\ }\textbf {\bibinfo {volume} {99}},\ \bibinfo
  {pages} {075125} (\bibinfo {year} {2019})}\BibitemShut {NoStop}%
\bibitem [{\citenamefont {Ament}\ \emph {et~al.}(2007)\citenamefont {Ament},
  \citenamefont {Forte},\ and\ \citenamefont {van~den Brink}}]{AmentPRB2007}%
  \BibitemOpen
  \bibfield  {author} {\bibinfo {author} {\bibfnamefont {L.~J.~P.}\
  \bibnamefont {Ament}}, \bibinfo {author} {\bibfnamefont {F.}~\bibnamefont
  {Forte}}, \ and\ \bibinfo {author} {\bibfnamefont {J.}~\bibnamefont {van~den
  Brink}},\ }\href {\doibase 10.1103/PhysRevB.75.115118} {\bibfield  {journal}
  {\bibinfo  {journal} {Phys. Rev. B}\ }\textbf {\bibinfo {volume} {75}},\
  \bibinfo {pages} {115118} (\bibinfo {year} {2007})}\BibitemShut {NoStop}%
\bibitem [{\citenamefont {Ament}\ \emph {et~al.}(2011)\citenamefont {Ament},
  \citenamefont {van Veenendaal}, \citenamefont {Devereaux}, \citenamefont
  {Hill},\ and\ \citenamefont {van~den Brink}}]{amentRMP2011}%
  \BibitemOpen
  \bibfield  {author} {\bibinfo {author} {\bibfnamefont {L.~J.~P.}\
  \bibnamefont {Ament}}, \bibinfo {author} {\bibfnamefont {M.}~\bibnamefont
  {van Veenendaal}}, \bibinfo {author} {\bibfnamefont {T.~P.}\ \bibnamefont
  {Devereaux}}, \bibinfo {author} {\bibfnamefont {J.~P.}\ \bibnamefont {Hill}},
  \ and\ \bibinfo {author} {\bibfnamefont {J.}~\bibnamefont {van~den Brink}},\
  }\href {\doibase 10.1103/RevModPhys.83.705} {\bibfield  {journal} {\bibinfo
  {journal} {Rev. Mod. Phys.}\ }\textbf {\bibinfo {volume} {83}},\ \bibinfo
  {pages} {705} (\bibinfo {year} {2011})}\BibitemShut {NoStop}%
\bibitem [{\citenamefont {Fukazawa}\ \emph {et~al.}(2000)\citenamefont
  {Fukazawa}, \citenamefont {Nakatsuji},\ and\ \citenamefont
  {Maeno}}]{FukazawaPhysB00}%
  \BibitemOpen
  \bibfield  {author} {\bibinfo {author} {\bibfnamefont {H.}~\bibnamefont
  {Fukazawa}}, \bibinfo {author} {\bibfnamefont {S.}~\bibnamefont {Nakatsuji}},
  \ and\ \bibinfo {author} {\bibfnamefont {Y.}~\bibnamefont {Maeno}},\ }\href
  {\doibase https://doi.org/10.1016/S0921-4526(99)00989-8} {\bibfield
  {journal} {\bibinfo  {journal} {Physica B}\ }\textbf {\bibinfo {volume}
  {281}},\ \bibinfo {pages} {613} (\bibinfo {year} {2000})}\BibitemShut
  {NoStop}%
\bibitem [{\citenamefont {Nakatsuji}\ and\ \citenamefont
  {Maeno}(2001)}]{snakatsujiJSSCHEM2001}%
  \BibitemOpen
  \bibfield  {author} {\bibinfo {author} {\bibfnamefont {S.}~\bibnamefont
  {Nakatsuji}}\ and\ \bibinfo {author} {\bibfnamefont {Y.}~\bibnamefont
  {Maeno}},\ }\href {\doibase 10.1006/jssc.2000.8953} {\bibfield  {journal}
  {\bibinfo  {journal} {Journal of Solid State Chemistry}\ }\textbf {\bibinfo
  {volume} {156}},\ \bibinfo {pages} {26 } (\bibinfo {year}
  {2001})}\BibitemShut {NoStop}%
\bibitem [{\citenamefont {Ghiringhelli}\ \emph {et~al.}(2006)\citenamefont
  {Ghiringhelli}, \citenamefont {Piazzalunga}, \citenamefont {Dallera},
  \citenamefont {Trezzi}, \citenamefont {Braicovich}, \citenamefont {Schmitt},
  \citenamefont {Strocov}, \citenamefont {Betemps}, \citenamefont {Patthey},
  \citenamefont {Wang},\ and\ \citenamefont
  {Grioni}}]{ghiringhelliREVSCIINS2006}%
  \BibitemOpen
  \bibfield  {author} {\bibinfo {author} {\bibfnamefont {G.}~\bibnamefont
  {Ghiringhelli}}, \bibinfo {author} {\bibfnamefont {A.}~\bibnamefont
  {Piazzalunga}}, \bibinfo {author} {\bibfnamefont {C.}~\bibnamefont
  {Dallera}}, \bibinfo {author} {\bibfnamefont {G.}~\bibnamefont {Trezzi}},
  \bibinfo {author} {\bibfnamefont {L.}~\bibnamefont {Braicovich}}, \bibinfo
  {author} {\bibfnamefont {T.}~\bibnamefont {Schmitt}}, \bibinfo {author}
  {\bibfnamefont {V.~N.}\ \bibnamefont {Strocov}}, \bibinfo {author}
  {\bibfnamefont {R.}~\bibnamefont {Betemps}}, \bibinfo {author} {\bibfnamefont
  {L.}~\bibnamefont {Patthey}}, \bibinfo {author} {\bibfnamefont
  {X.}~\bibnamefont {Wang}}, \ and\ \bibinfo {author} {\bibfnamefont
  {M.}~\bibnamefont {Grioni}},\ }\href {\doibase 10.1063/1.2372731} {\bibfield
  {journal} {\bibinfo  {journal} {Review of Scientific Instruments}\ }\textbf
  {\bibinfo {volume} {77}},\ \bibinfo {eid} {113108} (\bibinfo {year}
  {2006})}\BibitemShut {NoStop}%
\bibitem [{\citenamefont {Strocov}\ \emph {et~al.}(2010)\citenamefont
  {Strocov}, \citenamefont {Schmitt}, \citenamefont {Flechsig}, \citenamefont
  {Schmidt}, \citenamefont {Imhof}, \citenamefont {Chen}, \citenamefont
  {Raabe}, \citenamefont {Betemps}, \citenamefont {Zimoch}, \citenamefont
  {Krempasky} \emph {et~al.}}]{strocov2010high}%
  \BibitemOpen
  \bibfield  {author} {\bibinfo {author} {\bibfnamefont {V.}~\bibnamefont
  {Strocov}}, \bibinfo {author} {\bibfnamefont {T.}~\bibnamefont {Schmitt}},
  \bibinfo {author} {\bibfnamefont {U.}~\bibnamefont {Flechsig}}, \bibinfo
  {author} {\bibfnamefont {T.}~\bibnamefont {Schmidt}}, \bibinfo {author}
  {\bibfnamefont {A.}~\bibnamefont {Imhof}}, \bibinfo {author} {\bibfnamefont
  {Q.}~\bibnamefont {Chen}}, \bibinfo {author} {\bibfnamefont {J.}~\bibnamefont
  {Raabe}}, \bibinfo {author} {\bibfnamefont {R.}~\bibnamefont {Betemps}},
  \bibinfo {author} {\bibfnamefont {D.}~\bibnamefont {Zimoch}}, \bibinfo
  {author} {\bibfnamefont {J.}~\bibnamefont {Krempasky}},  \emph {et~al.},\
  }\href {\doibase 10.1107/S0909049510019862} {\bibfield  {journal} {\bibinfo
  {journal} {Journal of synchrotron radiation}\ }\textbf {\bibinfo {volume}
  {17}},\ \bibinfo {pages} {631} (\bibinfo {year} {2010})}\BibitemShut
  {NoStop}%
\bibitem [{\citenamefont {Chen}\ \emph {et~al.}(1992)\citenamefont {Chen},
  \citenamefont {Tjeng}, \citenamefont {Kwo}, \citenamefont {Kao},
  \citenamefont {Rudolf}, \citenamefont {Sette},\ and\ \citenamefont
  {Fleming}}]{chen_out--plane_1992}%
  \BibitemOpen
  \bibfield  {author} {\bibinfo {author} {\bibfnamefont {C.~T.}\ \bibnamefont
  {Chen}}, \bibinfo {author} {\bibfnamefont {L.~H.}\ \bibnamefont {Tjeng}},
  \bibinfo {author} {\bibfnamefont {J.}~\bibnamefont {Kwo}}, \bibinfo {author}
  {\bibfnamefont {H.~L.}\ \bibnamefont {Kao}}, \bibinfo {author} {\bibfnamefont
  {P.}~\bibnamefont {Rudolf}}, \bibinfo {author} {\bibfnamefont
  {F.}~\bibnamefont {Sette}}, \ and\ \bibinfo {author} {\bibfnamefont {R.~M.}\
  \bibnamefont {Fleming}},\ }\href {\doibase 10.1103/PhysRevLett.68.2543}
  {\bibfield  {journal} {\bibinfo  {journal} {Phys. Rev. Lett.}\ }\textbf
  {\bibinfo {volume} {68}},\ \bibinfo {pages} {2543} (\bibinfo {year}
  {1992})}\BibitemShut {NoStop}%
\bibitem [{\citenamefont {Noh}\ \emph {et~al.}(2005)\citenamefont {Noh},
  \citenamefont {Oh}, \citenamefont {Park}, \citenamefont {Park}, \citenamefont
  {Kim}, \citenamefont {Kim}, \citenamefont {Mizokawa}, \citenamefont {Tjeng},
  \citenamefont {Lin}, \citenamefont {Chen}, \citenamefont {Schuppler},
  \citenamefont {Nakatsuji}, \citenamefont {Fukazawa},\ and\ \citenamefont
  {Maeno}}]{NohPRB2005}%
  \BibitemOpen
  \bibfield  {author} {\bibinfo {author} {\bibfnamefont {H.-J.}\ \bibnamefont
  {Noh}}, \bibinfo {author} {\bibfnamefont {S.-J.}\ \bibnamefont {Oh}},
  \bibinfo {author} {\bibfnamefont {B.-G.}\ \bibnamefont {Park}}, \bibinfo
  {author} {\bibfnamefont {J.-H.}\ \bibnamefont {Park}}, \bibinfo {author}
  {\bibfnamefont {J.-Y.}\ \bibnamefont {Kim}}, \bibinfo {author} {\bibfnamefont
  {H.-D.}\ \bibnamefont {Kim}}, \bibinfo {author} {\bibfnamefont
  {T.}~\bibnamefont {Mizokawa}}, \bibinfo {author} {\bibfnamefont {L.~H.}\
  \bibnamefont {Tjeng}}, \bibinfo {author} {\bibfnamefont {H.-J.}\ \bibnamefont
  {Lin}}, \bibinfo {author} {\bibfnamefont {C.~T.}\ \bibnamefont {Chen}},
  \bibinfo {author} {\bibfnamefont {S.}~\bibnamefont {Schuppler}}, \bibinfo
  {author} {\bibfnamefont {S.}~\bibnamefont {Nakatsuji}}, \bibinfo {author}
  {\bibfnamefont {H.}~\bibnamefont {Fukazawa}}, \ and\ \bibinfo {author}
  {\bibfnamefont {Y.}~\bibnamefont {Maeno}},\ }\href {\doibase
  10.1103/PhysRevB.72.052411} {\bibfield  {journal} {\bibinfo  {journal} {Phys.
  Rev. B}\ }\textbf {\bibinfo {volume} {72}},\ \bibinfo {pages} {052411}
  (\bibinfo {year} {2005})}\BibitemShut {NoStop}%
\bibitem [{\citenamefont {Malvestuto}\ \emph {et~al.}(2011)\citenamefont
  {Malvestuto}, \citenamefont {Carleschi}, \citenamefont {Fittipaldi},
  \citenamefont {Gorelov}, \citenamefont {Pavarini}, \citenamefont {Cuoco},
  \citenamefont {Maeno}, \citenamefont {Parmigiani},\ and\ \citenamefont
  {Vecchione}}]{malvestutoPRB2011}%
  \BibitemOpen
  \bibfield  {author} {\bibinfo {author} {\bibfnamefont {M.}~\bibnamefont
  {Malvestuto}}, \bibinfo {author} {\bibfnamefont {E.}~\bibnamefont
  {Carleschi}}, \bibinfo {author} {\bibfnamefont {R.}~\bibnamefont
  {Fittipaldi}}, \bibinfo {author} {\bibfnamefont {E.}~\bibnamefont {Gorelov}},
  \bibinfo {author} {\bibfnamefont {E.}~\bibnamefont {Pavarini}}, \bibinfo
  {author} {\bibfnamefont {M.}~\bibnamefont {Cuoco}}, \bibinfo {author}
  {\bibfnamefont {Y.}~\bibnamefont {Maeno}}, \bibinfo {author} {\bibfnamefont
  {F.}~\bibnamefont {Parmigiani}}, \ and\ \bibinfo {author} {\bibfnamefont
  {A.}~\bibnamefont {Vecchione}},\ }\href {\doibase 10.1103/PhysRevB.83.165121}
  {\bibfield  {journal} {\bibinfo  {journal} {Phys. Rev. B}\ }\textbf {\bibinfo
  {volume} {83}},\ \bibinfo {pages} {165121} (\bibinfo {year}
  {2011})}\BibitemShut {NoStop}%
\bibitem [{\citenamefont {Malvestuto}\ \emph {et~al.}(2013)\citenamefont
  {Malvestuto}, \citenamefont {Capogrosso}, \citenamefont {Carleschi},
  \citenamefont {Galli}, \citenamefont {Gorelov}, \citenamefont {Pavarini},
  \citenamefont {Fittipaldi}, \citenamefont {Forte}, \citenamefont {Cuoco},
  \citenamefont {Vecchione},\ and\ \citenamefont
  {Parmigiani}}]{malvestutoPRB2013}%
  \BibitemOpen
  \bibfield  {author} {\bibinfo {author} {\bibfnamefont {M.}~\bibnamefont
  {Malvestuto}}, \bibinfo {author} {\bibfnamefont {V.}~\bibnamefont
  {Capogrosso}}, \bibinfo {author} {\bibfnamefont {E.}~\bibnamefont
  {Carleschi}}, \bibinfo {author} {\bibfnamefont {L.}~\bibnamefont {Galli}},
  \bibinfo {author} {\bibfnamefont {E.}~\bibnamefont {Gorelov}}, \bibinfo
  {author} {\bibfnamefont {E.}~\bibnamefont {Pavarini}}, \bibinfo {author}
  {\bibfnamefont {R.}~\bibnamefont {Fittipaldi}}, \bibinfo {author}
  {\bibfnamefont {F.}~\bibnamefont {Forte}}, \bibinfo {author} {\bibfnamefont
  {M.}~\bibnamefont {Cuoco}}, \bibinfo {author} {\bibfnamefont
  {A.}~\bibnamefont {Vecchione}}, \ and\ \bibinfo {author} {\bibfnamefont
  {F.}~\bibnamefont {Parmigiani}},\ }\href {\doibase
  https://doi.org/10.1103/PhysRevB.88.195143} {\bibfield  {journal} {\bibinfo
  {journal} {Phys. Rev. B}\ }\textbf {\bibinfo {volume} {88}},\ \bibinfo
  {pages} {195143} (\bibinfo {year} {2013})}\BibitemShut {NoStop}%
\bibitem [{\citenamefont {Guedes}\ \emph {et~al.}(2012)\citenamefont {Guedes},
  \citenamefont {Abbate}, \citenamefont {Ishigami}, \citenamefont {Fujimori},
  \citenamefont {Yoshimatsu}, \citenamefont {Kumigashira}, \citenamefont
  {Oshima}, \citenamefont {Vicentin}, \citenamefont {Fonseca},\ and\
  \citenamefont {Mossanek}}]{guedes_core_2012}%
  \BibitemOpen
  \bibfield  {author} {\bibinfo {author} {\bibfnamefont {E.~B.}\ \bibnamefont
  {Guedes}}, \bibinfo {author} {\bibfnamefont {M.}~\bibnamefont {Abbate}},
  \bibinfo {author} {\bibfnamefont {K.}~\bibnamefont {Ishigami}}, \bibinfo
  {author} {\bibfnamefont {A.}~\bibnamefont {Fujimori}}, \bibinfo {author}
  {\bibfnamefont {K.}~\bibnamefont {Yoshimatsu}}, \bibinfo {author}
  {\bibfnamefont {H.}~\bibnamefont {Kumigashira}}, \bibinfo {author}
  {\bibfnamefont {M.}~\bibnamefont {Oshima}}, \bibinfo {author} {\bibfnamefont
  {F.~C.}\ \bibnamefont {Vicentin}}, \bibinfo {author} {\bibfnamefont {P.~T.}\
  \bibnamefont {Fonseca}}, \ and\ \bibinfo {author} {\bibfnamefont {R.~J.~O.}\
  \bibnamefont {Mossanek}},\ }\href {\doibase 10.1103/PhysRevB.86.235127}
  {\bibfield  {journal} {\bibinfo  {journal} {Phys. Rev. B}\ }\textbf {\bibinfo
  {volume} {86}},\ \bibinfo {pages} {235127} (\bibinfo {year}
  {2012})}\BibitemShut {NoStop}%
\bibitem [{\citenamefont {Koga}\ \emph {et~al.}(2004)\citenamefont {Koga},
  \citenamefont {Kawakami}, \citenamefont {Rice},\ and\ \citenamefont
  {Sigrist}}]{KogaPRL2004}%
  \BibitemOpen
  \bibfield  {author} {\bibinfo {author} {\bibfnamefont {A.}~\bibnamefont
  {Koga}}, \bibinfo {author} {\bibfnamefont {N.}~\bibnamefont {Kawakami}},
  \bibinfo {author} {\bibfnamefont {T.~M.}\ \bibnamefont {Rice}}, \ and\
  \bibinfo {author} {\bibfnamefont {M.}~\bibnamefont {Sigrist}},\ }\href
  {\doibase 10.1103/PhysRevLett.92.216402} {\bibfield  {journal} {\bibinfo
  {journal} {Phys. Rev. Lett.}\ }\textbf {\bibinfo {volume} {92}},\ \bibinfo
  {pages} {216402} (\bibinfo {year} {2004})}\BibitemShut {NoStop}%
\bibitem [{\citenamefont {Liebsch}\ and\ \citenamefont
  {Ishida}(2007)}]{LiebschPRL2007}%
  \BibitemOpen
  \bibfield  {author} {\bibinfo {author} {\bibfnamefont {A.}~\bibnamefont
  {Liebsch}}\ and\ \bibinfo {author} {\bibfnamefont {H.}~\bibnamefont
  {Ishida}},\ }\href {\doibase 10.1103/PhysRevLett.98.216403} {\bibfield
  {journal} {\bibinfo  {journal} {Phys. Rev. Lett.}\ }\textbf {\bibinfo
  {volume} {98}},\ \bibinfo {pages} {216403} (\bibinfo {year}
  {2007})}\BibitemShut {NoStop}%
\bibitem [{\citenamefont {Gorelov}\ \emph {et~al.}(2010)\citenamefont
  {Gorelov}, \citenamefont {Karolak}, \citenamefont {Wehling}, \citenamefont
  {Lechermann}, \citenamefont {Lichtenstein},\ and\ \citenamefont
  {Pavarini}}]{gorelovPRL2010}%
  \BibitemOpen
  \bibfield  {author} {\bibinfo {author} {\bibfnamefont {E.}~\bibnamefont
  {Gorelov}}, \bibinfo {author} {\bibfnamefont {M.}~\bibnamefont {Karolak}},
  \bibinfo {author} {\bibfnamefont {T.~O.}\ \bibnamefont {Wehling}}, \bibinfo
  {author} {\bibfnamefont {F.}~\bibnamefont {Lechermann}}, \bibinfo {author}
  {\bibfnamefont {A.~I.}\ \bibnamefont {Lichtenstein}}, \ and\ \bibinfo
  {author} {\bibfnamefont {E.}~\bibnamefont {Pavarini}},\ }\href {\doibase
  10.1103/PhysRevLett.104.226401} {\bibfield  {journal} {\bibinfo  {journal}
  {Phys. Rev. Lett.}\ }\textbf {\bibinfo {volume} {104}},\ \bibinfo {pages}
  {226401} (\bibinfo {year} {2010})}\BibitemShut {NoStop}%
\bibitem [{\citenamefont {Braden}\ \emph {et~al.}(2007)\citenamefont {Braden},
  \citenamefont {Reichardt}, \citenamefont {Sidis}, \citenamefont {Mao},\ and\
  \citenamefont {Maeno}}]{braden_lattice_nodate}%
  \BibitemOpen
  \bibfield  {author} {\bibinfo {author} {\bibfnamefont {M.}~\bibnamefont
  {Braden}}, \bibinfo {author} {\bibfnamefont {W.}~\bibnamefont {Reichardt}},
  \bibinfo {author} {\bibfnamefont {Y.}~\bibnamefont {Sidis}}, \bibinfo
  {author} {\bibfnamefont {Z.}~\bibnamefont {Mao}}, \ and\ \bibinfo {author}
  {\bibfnamefont {Y.}~\bibnamefont {Maeno}},\ }\href {\doibase
  10.1103/PhysRevB.76.014505} {\bibfield  {journal} {\bibinfo  {journal} {Phys.
  Rev. B}\ }\textbf {\bibinfo {volume} {76}},\ \bibinfo {pages} {014505}
  (\bibinfo {year} {2007})}\BibitemShut {NoStop}%
\bibitem [{\citenamefont {Gretarsson}\ \emph {et~al.}(2019)\citenamefont
  {Gretarsson}, \citenamefont {Suzuki}, \citenamefont {Kim}, \citenamefont
  {Ueda}, \citenamefont {Krautloher}, \citenamefont {Kim}, \citenamefont
  {Yava\ifmmode~\mbox{\c{s}}\else \c{s}\fi{}}, \citenamefont {Khaliullin},\
  and\ \citenamefont {Keimer}}]{GretarssonPRB2019}%
  \BibitemOpen
  \bibfield  {author} {\bibinfo {author} {\bibfnamefont {H.}~\bibnamefont
  {Gretarsson}}, \bibinfo {author} {\bibfnamefont {H.}~\bibnamefont {Suzuki}},
  \bibinfo {author} {\bibfnamefont {H.}~\bibnamefont {Kim}}, \bibinfo {author}
  {\bibfnamefont {K.}~\bibnamefont {Ueda}}, \bibinfo {author} {\bibfnamefont
  {M.}~\bibnamefont {Krautloher}}, \bibinfo {author} {\bibfnamefont {B.~J.}\
  \bibnamefont {Kim}}, \bibinfo {author} {\bibfnamefont {H.}~\bibnamefont
  {Yava\ifmmode~\mbox{\c{s}}\else \c{s}\fi{}}}, \bibinfo {author}
  {\bibfnamefont {G.}~\bibnamefont {Khaliullin}}, \ and\ \bibinfo {author}
  {\bibfnamefont {B.}~\bibnamefont {Keimer}},\ }\href {\doibase
  10.1103/PhysRevB.100.045123} {\bibfield  {journal} {\bibinfo  {journal}
  {Phys. Rev. B}\ }\textbf {\bibinfo {volume} {100}},\ \bibinfo {pages}
  {045123} (\bibinfo {year} {2019})}\BibitemShut {NoStop}%
\bibitem [{\citenamefont {Mizokawa}\ \emph {et~al.}(2001)\citenamefont
  {Mizokawa}, \citenamefont {Tjeng}, \citenamefont {Sawatzky}, \citenamefont
  {Ghiringhelli}, \citenamefont {Tjernberg}, \citenamefont {Brookes},
  \citenamefont {Fukazawa}, \citenamefont {Nakatsuji},\ and\ \citenamefont
  {Maeno}}]{mizokawaPRL2001}%
  \BibitemOpen
  \bibfield  {author} {\bibinfo {author} {\bibfnamefont {T.}~\bibnamefont
  {Mizokawa}}, \bibinfo {author} {\bibfnamefont {L.~H.}\ \bibnamefont {Tjeng}},
  \bibinfo {author} {\bibfnamefont {G.~A.}\ \bibnamefont {Sawatzky}}, \bibinfo
  {author} {\bibfnamefont {G.}~\bibnamefont {Ghiringhelli}}, \bibinfo {author}
  {\bibfnamefont {O.}~\bibnamefont {Tjernberg}}, \bibinfo {author}
  {\bibfnamefont {N.~B.}\ \bibnamefont {Brookes}}, \bibinfo {author}
  {\bibfnamefont {H.}~\bibnamefont {Fukazawa}}, \bibinfo {author}
  {\bibfnamefont {S.}~\bibnamefont {Nakatsuji}}, \ and\ \bibinfo {author}
  {\bibfnamefont {Y.}~\bibnamefont {Maeno}},\ }\href {\doibase
  10.1103/PhysRevLett.87.077202} {\bibfield  {journal} {\bibinfo  {journal}
  {Phys. Rev. Lett.}\ }\textbf {\bibinfo {volume} {87}},\ \bibinfo {pages}
  {077202} (\bibinfo {year} {2001})}\BibitemShut {NoStop}%
\bibitem [{\citenamefont {Veenstra}\ \emph {et~al.}(2014)\citenamefont
  {Veenstra}, \citenamefont {Zhu}, \citenamefont {Raichle}, \citenamefont
  {Ludbrook}, \citenamefont {Nicolaou}, \citenamefont {Slomski}, \citenamefont
  {Landolt}, \citenamefont {Kittaka}, \citenamefont {Maeno}, \citenamefont
  {Dil}, \citenamefont {Elfimov}, \citenamefont {Haverkort},\ and\
  \citenamefont {Damascelli}}]{veenstraPRL2014}%
  \BibitemOpen
  \bibfield  {author} {\bibinfo {author} {\bibfnamefont {C.~N.}\ \bibnamefont
  {Veenstra}}, \bibinfo {author} {\bibfnamefont {Z.-H.}\ \bibnamefont {Zhu}},
  \bibinfo {author} {\bibfnamefont {M.}~\bibnamefont {Raichle}}, \bibinfo
  {author} {\bibfnamefont {B.~M.}\ \bibnamefont {Ludbrook}}, \bibinfo {author}
  {\bibfnamefont {A.}~\bibnamefont {Nicolaou}}, \bibinfo {author}
  {\bibfnamefont {B.}~\bibnamefont {Slomski}}, \bibinfo {author} {\bibfnamefont
  {G.}~\bibnamefont {Landolt}}, \bibinfo {author} {\bibfnamefont
  {S.}~\bibnamefont {Kittaka}}, \bibinfo {author} {\bibfnamefont
  {Y.}~\bibnamefont {Maeno}}, \bibinfo {author} {\bibfnamefont {J.~H.}\
  \bibnamefont {Dil}}, \bibinfo {author} {\bibfnamefont {I.~S.}\ \bibnamefont
  {Elfimov}}, \bibinfo {author} {\bibfnamefont {M.~W.}\ \bibnamefont
  {Haverkort}}, \ and\ \bibinfo {author} {\bibfnamefont {A.}~\bibnamefont
  {Damascelli}},\ }\href {\doibase 10.1103/PhysRevLett.112.127002} {\bibfield
  {journal} {\bibinfo  {journal} {Phys. Rev. Lett.}\ }\textbf {\bibinfo
  {volume} {112}},\ \bibinfo {pages} {127002} (\bibinfo {year}
  {2014})}\BibitemShut {NoStop}%
\bibitem [{\citenamefont {Han}\ and\ \citenamefont
  {Millis}(2018)}]{han_lattice_2018}%
  \BibitemOpen
  \bibfield  {author} {\bibinfo {author} {\bibfnamefont {Q.}~\bibnamefont
  {Han}}\ and\ \bibinfo {author} {\bibfnamefont {A.}~\bibnamefont {Millis}},\
  }\href {\doibase 10.1103/PhysRevLett.121.067601} {\bibfield  {journal}
  {\bibinfo  {journal} {Phys. Rev. Lett.}\ }\textbf {\bibinfo {volume} {121}},\
  \bibinfo {pages} {067601} (\bibinfo {year} {2018})}\BibitemShut {NoStop}%
\bibitem [{\citenamefont {Sutter}\ \emph {et~al.}(2017)\citenamefont {Sutter},
  \citenamefont {Fatuzzo}, \citenamefont {Moser}, \citenamefont {Kim},
  \citenamefont {Fittipaldi}, \citenamefont {Vecchione}, \citenamefont
  {Granata}, \citenamefont {Sassa}, \citenamefont {Cossalter}, \citenamefont
  {Gatti}, \citenamefont {Grioni}, \citenamefont {R\o{}nnow}, \citenamefont
  {Plumb}, \citenamefont {Matt}, \citenamefont {Shi}, \citenamefont {Hoesch},
  \citenamefont {Kim}, \citenamefont {Chang}, \citenamefont {Jeng},
  \citenamefont {Jozwiak}, \citenamefont {Bostwick}, \citenamefont {Rotenberg},
  \citenamefont {Georges}, \citenamefont {Neupert},\ and\ \citenamefont
  {Chang}}]{SutterNatComm2017a}%
  \BibitemOpen
  \bibfield  {author} {\bibinfo {author} {\bibfnamefont {D.}~\bibnamefont
  {Sutter}}, \bibinfo {author} {\bibfnamefont {C.}~\bibnamefont {Fatuzzo}},
  \bibinfo {author} {\bibfnamefont {S.}~\bibnamefont {Moser}}, \bibinfo
  {author} {\bibfnamefont {M.}~\bibnamefont {Kim}}, \bibinfo {author}
  {\bibfnamefont {R.}~\bibnamefont {Fittipaldi}}, \bibinfo {author}
  {\bibfnamefont {A.}~\bibnamefont {Vecchione}}, \bibinfo {author}
  {\bibfnamefont {V.}~\bibnamefont {Granata}}, \bibinfo {author} {\bibfnamefont
  {Y.}~\bibnamefont {Sassa}}, \bibinfo {author} {\bibfnamefont
  {F.}~\bibnamefont {Cossalter}}, \bibinfo {author} {\bibfnamefont
  {G.}~\bibnamefont {Gatti}}, \bibinfo {author} {\bibfnamefont
  {M.}~\bibnamefont {Grioni}}, \bibinfo {author} {\bibfnamefont {H.~M.}\
  \bibnamefont {R\o{}nnow}}, \bibinfo {author} {\bibfnamefont {N.~C.}\
  \bibnamefont {Plumb}}, \bibinfo {author} {\bibfnamefont {C.~E.}\ \bibnamefont
  {Matt}}, \bibinfo {author} {\bibfnamefont {M.}~\bibnamefont {Shi}}, \bibinfo
  {author} {\bibfnamefont {M.}~\bibnamefont {Hoesch}}, \bibinfo {author}
  {\bibfnamefont {T.~K.}\ \bibnamefont {Kim}}, \bibinfo {author} {\bibfnamefont
  {T.~R.}\ \bibnamefont {Chang}}, \bibinfo {author} {\bibfnamefont {H.~T.}\
  \bibnamefont {Jeng}}, \bibinfo {author} {\bibfnamefont {C.}~\bibnamefont
  {Jozwiak}}, \bibinfo {author} {\bibfnamefont {A.}~\bibnamefont {Bostwick}},
  \bibinfo {author} {\bibfnamefont {E.}~\bibnamefont {Rotenberg}}, \bibinfo
  {author} {\bibfnamefont {A.}~\bibnamefont {Georges}}, \bibinfo {author}
  {\bibfnamefont {T.}~\bibnamefont {Neupert}}, \ and\ \bibinfo {author}
  {\bibfnamefont {J.}~\bibnamefont {Chang}},\ }\href {\doibase
  10.1038/ncomms15176} {\bibfield  {journal} {\bibinfo  {journal} {Nat. Comm.}\
  }\textbf {\bibinfo {volume} {8}},\ \bibinfo {pages} {15176} (\bibinfo {year}
  {2017})}\BibitemShut {NoStop}%
\bibitem [{\citenamefont {Porter}\ \emph {et~al.}(2018)\citenamefont {Porter},
  \citenamefont {Granata}, \citenamefont {Forte}, \citenamefont {Di~Matteo},
  \citenamefont {Cuoco}, \citenamefont {Fittipaldi}, \citenamefont
  {Vecchione},\ and\ \citenamefont {Bombardi}}]{porterPRB2018}%
  \BibitemOpen
  \bibfield  {author} {\bibinfo {author} {\bibfnamefont {D.~G.}\ \bibnamefont
  {Porter}}, \bibinfo {author} {\bibfnamefont {V.}~\bibnamefont {Granata}},
  \bibinfo {author} {\bibfnamefont {F.}~\bibnamefont {Forte}}, \bibinfo
  {author} {\bibfnamefont {S.}~\bibnamefont {Di~Matteo}}, \bibinfo {author}
  {\bibfnamefont {M.}~\bibnamefont {Cuoco}}, \bibinfo {author} {\bibfnamefont
  {R.}~\bibnamefont {Fittipaldi}}, \bibinfo {author} {\bibfnamefont
  {A.}~\bibnamefont {Vecchione}}, \ and\ \bibinfo {author} {\bibfnamefont
  {A.}~\bibnamefont {Bombardi}},\ }\href {\doibase
  https://doi.org/10.1103/PhysRevB.98.125142} {\bibfield  {journal} {\bibinfo
  {journal} {Phys. Rev. B}\ }\textbf {\bibinfo {volume} {98}},\ \bibinfo
  {pages} {125142} (\bibinfo {year} {2018})}\BibitemShut {NoStop}%
\bibitem [{\citenamefont {Wang}\ \emph {et~al.}(2010)\citenamefont {Wang},
  \citenamefont {Wang}, \citenamefont {Saal}, \citenamefont {Shang},
  \citenamefont {Chen},\ and\ \citenamefont {Liu}}]{wang_phonon_nodate}%
  \BibitemOpen
  \bibfield  {author} {\bibinfo {author} {\bibfnamefont {Y.}~\bibnamefont
  {Wang}}, \bibinfo {author} {\bibfnamefont {J.~J.}\ \bibnamefont {Wang}},
  \bibinfo {author} {\bibfnamefont {J.~E.}\ \bibnamefont {Saal}}, \bibinfo
  {author} {\bibfnamefont {S.~L.}\ \bibnamefont {Shang}}, \bibinfo {author}
  {\bibfnamefont {L.-Q.}\ \bibnamefont {Chen}}, \ and\ \bibinfo {author}
  {\bibfnamefont {Z.-K.}\ \bibnamefont {Liu}},\ }\href {\doibase
  10.1103/PhysRevB.82.172503} {\bibfield  {journal} {\bibinfo  {journal} {Phys.
  Rev. B}\ }\textbf {\bibinfo {volume} {82}},\ \bibinfo {pages} {172503}
  (\bibinfo {year} {2010})}\BibitemShut {NoStop}%
\bibitem [{\citenamefont {Dean}\ \emph {et~al.}(2013)\citenamefont {Dean},
  \citenamefont {Dellea}, \citenamefont {Springell}, \citenamefont
  {Yakhou-Harris}, \citenamefont {Kummer}, \citenamefont {Brookes},
  \citenamefont {Liu}, \citenamefont {Sun}, \citenamefont {Strle},
  \citenamefont {Schmitt}, \citenamefont {Braicovich}, \citenamefont
  {Ghiringhelli}, \citenamefont {Bo{\v z}ovi{\'c}},\ and\ \citenamefont
  {Hill}}]{DeanNatMat2013}%
  \BibitemOpen
  \bibfield  {author} {\bibinfo {author} {\bibfnamefont {M.~P.~M.}\
  \bibnamefont {Dean}}, \bibinfo {author} {\bibfnamefont {G.}~\bibnamefont
  {Dellea}}, \bibinfo {author} {\bibfnamefont {R.~S.}\ \bibnamefont
  {Springell}}, \bibinfo {author} {\bibfnamefont {F.}~\bibnamefont
  {Yakhou-Harris}}, \bibinfo {author} {\bibfnamefont {K.}~\bibnamefont
  {Kummer}}, \bibinfo {author} {\bibfnamefont {N.~B.}\ \bibnamefont {Brookes}},
  \bibinfo {author} {\bibfnamefont {X.}~\bibnamefont {Liu}}, \bibinfo {author}
  {\bibfnamefont {Y.-J.}\ \bibnamefont {Sun}}, \bibinfo {author} {\bibfnamefont
  {J.}~\bibnamefont {Strle}}, \bibinfo {author} {\bibfnamefont
  {T.}~\bibnamefont {Schmitt}}, \bibinfo {author} {\bibfnamefont
  {L.}~\bibnamefont {Braicovich}}, \bibinfo {author} {\bibfnamefont
  {G.}~\bibnamefont {Ghiringhelli}}, \bibinfo {author} {\bibfnamefont
  {I.}~\bibnamefont {Bo{\v z}ovi{\'c}}}, \ and\ \bibinfo {author}
  {\bibfnamefont {J.~P.}\ \bibnamefont {Hill}},\ }\href
  {https://doi.org/10.1038/nmat3723} {\bibfield  {journal} {\bibinfo  {journal}
  {Nature Materials}\ }\textbf {\bibinfo {volume} {12}},\ \bibinfo {pages}
  {1019 EP } (\bibinfo {year} {2013})}\BibitemShut {NoStop}%
\bibitem [{\citenamefont {Monney}\ \emph {et~al.}(2016)\citenamefont {Monney},
  \citenamefont {Schmitt}, \citenamefont {Matt}, \citenamefont {Mesot},
  \citenamefont {Strocov}, \citenamefont {Lipscombe}, \citenamefont {Hayden},\
  and\ \citenamefont {Chang}}]{monney_resonant_2016}%
  \BibitemOpen
  \bibfield  {author} {\bibinfo {author} {\bibfnamefont {C.}~\bibnamefont
  {Monney}}, \bibinfo {author} {\bibfnamefont {T.}~\bibnamefont {Schmitt}},
  \bibinfo {author} {\bibfnamefont {C.~E.}\ \bibnamefont {Matt}}, \bibinfo
  {author} {\bibfnamefont {J.}~\bibnamefont {Mesot}}, \bibinfo {author}
  {\bibfnamefont {V.~N.}\ \bibnamefont {Strocov}}, \bibinfo {author}
  {\bibfnamefont {O.~J.}\ \bibnamefont {Lipscombe}}, \bibinfo {author}
  {\bibfnamefont {S.~M.}\ \bibnamefont {Hayden}}, \ and\ \bibinfo {author}
  {\bibfnamefont {J.}~\bibnamefont {Chang}},\ }\href {\doibase
  10.1103/PhysRevB.93.075103} {\bibfield  {journal} {\bibinfo  {journal} {Phys.
  Rev. B}\ }\textbf {\bibinfo {volume} {93}},\ \bibinfo {pages} {075103}
  (\bibinfo {year} {2016})}\BibitemShut {NoStop}%
\bibitem [{\citenamefont {Pelliciari}\ \emph {et~al.}(2019)\citenamefont
  {Pelliciari}, \citenamefont {Ishii}, \citenamefont {Huang}, \citenamefont
  {Dantz}, \citenamefont {Lu}, \citenamefont {Olalde-Velasco}, \citenamefont
  {Strocov}, \citenamefont {Kasahara}, \citenamefont {Xing}, \citenamefont
  {Wang}, \citenamefont {Jin}, \citenamefont {Matsuda}, \citenamefont
  {Shibauchi}, \citenamefont {Das},\ and\ \citenamefont
  {Schmitt}}]{pelliciariComPhys2019}%
  \BibitemOpen
  \bibfield  {author} {\bibinfo {author} {\bibfnamefont {J.}~\bibnamefont
  {Pelliciari}}, \bibinfo {author} {\bibfnamefont {K.}~\bibnamefont {Ishii}},
  \bibinfo {author} {\bibfnamefont {Y.}~\bibnamefont {Huang}}, \bibinfo
  {author} {\bibfnamefont {M.}~\bibnamefont {Dantz}}, \bibinfo {author}
  {\bibfnamefont {X.}~\bibnamefont {Lu}}, \bibinfo {author} {\bibfnamefont
  {P.}~\bibnamefont {Olalde-Velasco}}, \bibinfo {author} {\bibfnamefont
  {V.~N.}\ \bibnamefont {Strocov}}, \bibinfo {author} {\bibfnamefont
  {S.}~\bibnamefont {Kasahara}}, \bibinfo {author} {\bibfnamefont
  {L.}~\bibnamefont {Xing}}, \bibinfo {author} {\bibfnamefont {X.}~\bibnamefont
  {Wang}}, \bibinfo {author} {\bibfnamefont {C.}~\bibnamefont {Jin}}, \bibinfo
  {author} {\bibfnamefont {Y.}~\bibnamefont {Matsuda}}, \bibinfo {author}
  {\bibfnamefont {T.}~\bibnamefont {Shibauchi}}, \bibinfo {author}
  {\bibfnamefont {T.}~\bibnamefont {Das}}, \ and\ \bibinfo {author}
  {\bibfnamefont {T.}~\bibnamefont {Schmitt}},\ }\href {\doibase
  10.1038/s42005-019-0236-3} {\bibfield  {journal} {\bibinfo  {journal}
  {Communications Physics}\ }\textbf {\bibinfo {volume} {2}},\ \bibinfo {pages}
  {139} (\bibinfo {year} {2019})}\BibitemShut {NoStop}%
\bibitem [{\citenamefont {Forte}\ \emph {et~al.}(2010)\citenamefont {Forte},
  \citenamefont {Cuoco},\ and\ \citenamefont {Noce}}]{fortePRB2010}%
  \BibitemOpen
  \bibfield  {author} {\bibinfo {author} {\bibfnamefont {F.}~\bibnamefont
  {Forte}}, \bibinfo {author} {\bibfnamefont {M.}~\bibnamefont {Cuoco}}, \ and\
  \bibinfo {author} {\bibfnamefont {C.}~\bibnamefont {Noce}},\ }\href {\doibase
  https://doi.org/10.1103/PhysRevB.82.155104} {\bibfield  {journal} {\bibinfo
  {journal} {Phys. Rev. B}\ }\textbf {\bibinfo {volume} {82}},\ \bibinfo
  {pages} {155104} (\bibinfo {year} {2010})}\BibitemShut {NoStop}%
\bibitem [{\citenamefont {Cuoco}\ \emph
  {et~al.}(2006{\natexlab{a}})\citenamefont {Cuoco}, \citenamefont {Forte},\
  and\ \citenamefont {Noce}}]{CuocoPRB06a}%
  \BibitemOpen
  \bibfield  {author} {\bibinfo {author} {\bibfnamefont {M.}~\bibnamefont
  {Cuoco}}, \bibinfo {author} {\bibfnamefont {F.}~\bibnamefont {Forte}}, \ and\
  \bibinfo {author} {\bibfnamefont {C.}~\bibnamefont {Noce}},\ }\href@noop {}
  {\bibfield  {journal} {\bibinfo  {journal} {Phys. Rev. B}\ }\textbf {\bibinfo
  {volume} {74}},\ \bibinfo {pages} {195124} (\bibinfo {year}
  {2006}{\natexlab{a}})}\BibitemShut {NoStop}%
\bibitem [{\citenamefont {Cuoco}\ \emph
  {et~al.}(2006{\natexlab{b}})\citenamefont {Cuoco}, \citenamefont {Forte},\
  and\ \citenamefont {Noce}}]{CuocoPRB06b}%
  \BibitemOpen
  \bibfield  {author} {\bibinfo {author} {\bibfnamefont {M.}~\bibnamefont
  {Cuoco}}, \bibinfo {author} {\bibfnamefont {F.}~\bibnamefont {Forte}}, \ and\
  \bibinfo {author} {\bibfnamefont {C.}~\bibnamefont {Noce}},\ }\href@noop {}
  {\bibfield  {journal} {\bibinfo  {journal} {Phys. Rev. B}\ }\textbf {\bibinfo
  {volume} {73}},\ \bibinfo {pages} {094428} (\bibinfo {year}
  {2006}{\natexlab{b}})}\BibitemShut {NoStop}%
\bibitem [{\citenamefont {Slater}\ and\ \citenamefont
  {Koster}(1954)}]{slaterPR1954}%
  \BibitemOpen
  \bibfield  {author} {\bibinfo {author} {\bibfnamefont {J.~C.}\ \bibnamefont
  {Slater}}\ and\ \bibinfo {author} {\bibfnamefont {G.~F.}\ \bibnamefont
  {Koster}},\ }\href {\doibase 10.1103/PhysRev.94.1498} {\bibfield  {journal}
  {\bibinfo  {journal} {Physical Review}\ }\textbf {\bibinfo {volume} {94}},\
  \bibinfo {pages} {1498} (\bibinfo {year} {1954})}\BibitemShut {NoStop}%
\bibitem [{\citenamefont {Brzezicki}\ \emph {et~al.}(2015)\citenamefont
  {Brzezicki}, \citenamefont {Noce}, \citenamefont {Romano},\ and\
  \citenamefont {Cuoco}}]{brzezickiPRL2015}%
  \BibitemOpen
  \bibfield  {author} {\bibinfo {author} {\bibfnamefont {W.}~\bibnamefont
  {Brzezicki}}, \bibinfo {author} {\bibfnamefont {C.}~\bibnamefont {Noce}},
  \bibinfo {author} {\bibfnamefont {A.}~\bibnamefont {Romano}}, \ and\ \bibinfo
  {author} {\bibfnamefont {M.}~\bibnamefont {Cuoco}},\ }\href {\doibase
  https://doi.org/10.1103/PhysRevLett.114.247002} {\bibfield  {journal}
  {\bibinfo  {journal} {Phys. Rev. Lett.}\ }\textbf {\bibinfo {volume} {114}},\
  \bibinfo {pages} {247002} (\bibinfo {year} {2015})}\BibitemShut {NoStop}%
\end{thebibliography}%

\end{document}